\documentclass{article}[12pt,a4paper]
\usepackage{graphicx}  
\usepackage{dcolumn}   
\usepackage{bm}        
\usepackage{amssymb}   
\usepackage[]{amsmath,amssymb}
\usepackage{graphics,epsfig}
\usepackage[latin1]{inputenc}
\usepackage[T1]{fontenc}
\usepackage{amsfonts}
\usepackage{subcaption} 
\usepackage{latexsym}
\usepackage[english]{babel}
\newcommand{\be}{\begin{equation}}
\newcommand{\ee}{\end{equation}}
\newcommand{\beq}{\begin{equation}}
\newcommand{\eeq}{\end{equation}}
\newcommand{\bea}{\begin{eqnarray}}
\newcommand{\eea}{\end{eqnarray}}
\newcommand{\nn}{\nonumber}
\def\be{\begin{equation}}
\def\ee{\end{equation}}
\def\ba{\begin{eqnarray}}
\def\ea{\end{eqnarray}}

\begin{document}


\title{Local temperatures and local terms in modular Hamiltonians}

\author{Ra\'{u}l E. Arias$^{a,b}$, David D. Blanco$^c$, Horacio Casini$^b$, Marina Huerta$^b$}

\maketitle

\begin{center}

{\sl $^a$ Instituto de F\'{\i}sica de La Plata - CONICET\\
C.C. 67, 1900 La Plata, Argentina.}

~

{\sl $^b$ Centro At\'omico Bariloche,\\
8400-S.C. de Bariloche, R\'{\i}o Negro, Argentina}

~

{\sl $^c$ CONICET - Universidad de Buenos Aires. \\Instituto de Astronom\'{\i}a y F\'{\i}sica del
Espacio (IAFE). \\Buenos Aires, Argentina.}

\end{center}

~

~

\begin{abstract}
We show there are analogues to the Unruh temperature that can be defined for any quantum field theory and region of the space. These local temperatures are defined using relative entropy with localized excitations. We show important restrictions arise from relative entropy inequalities and causal propagation between Cauchy surfaces. These suggest a large amount of universality for local temperatures, specially the ones affecting null directions. For regions with any number of intervals in two space-time dimensions the local temperatures might arise from a term in the modular Hamiltonian proportional to the stress tensor. We argue this term might be universal, with a coefficient that is the same for any theory, and check analytically and numerically this is the case for free massive scalar and Dirac fields. In dimensions $d\ge 3$ the local terms in the modular Hamiltonian producing these local temperatures cannot be formed exclusively from the stress tensor. For a free scalar field we classify the structure of the local terms.
\end{abstract}

\newpage

\tableofcontents

\section{Introduction}
In relativistic quantum field theory (QFT) the reduced density matrix corresponding to the vacuum state on the half spatial plane $x^0=0, x^1>0$ is given by a universal expression in terms of the stress tensor
\be
\rho= c\, e^{-2 \pi \int_{x^1>0} d^{d-1}x\, x^1 \, T_{00}(x) }\,.\label{f1}
\ee
This important result follows at the axiomatic level from analyticity properties originating in Lorentz invariance and positivity of energy, and is tightly related to the CPT theorem \cite{bisognano}. In the path integral Euclidean formulation is simply due to the fact that the density matrix in half space has a representation in terms of a $2\pi$ rotation in the $(x^0,x^1)$ plane (see for example \cite{example,exa1}).

Even if (\ref{f1}) has a fairly simple derivation, its physical interpretation has deservedly caused some wonder along time. First, it tells about an intriguing relation between entanglement in vacuum and energy density. This relation is behind the validity of entropy bounds coming from black hole physics in the weak gravity limit, namely Bekenstein's bound \cite{bekenstein}, the generalized second law \cite{gsl}, and the Bousso bound \cite{bousso}.

Second, if we write a density matrix as $\rho=e^{-K}$, where $K$ is called the modular Hamiltonian, eq. (\ref{f1}) tells the modular Hamiltonian for half space is an integral of a local operator. $K$ is in fact $2\pi$ times the generator of boosts. For a conformal field theory (CFT) $K$ is also local for spheres. The locality is then related to a Killing symmetry of the Rindler Wedge\footnote{The Rindler wedge is the space-time region $x^1>|x^0|$ which is the domain of dependence or causal completion of the half spatial plane $x^0=0, x^1>0$.}, and in the case of spheres in a CFT, to a conformal Killing symmetry.  However, from the point of view of quantum information theory this is rather mysterious.  On general grounds we do not expect locality to hold for non relativistic theories, or for the reduced density matrices of regions different from half space or non vacuum states; in general $K$ will be given by a non local and non linear combination of the field operators inside the region.

Formula (\ref{f1}) is related to Unruh temperature for accelerated observers \cite{unruh} (and to Hawking temperature of black holes). These observers evolve in time along boost orbits, and hence for them, time translations in the Rindler wedge are generated by the modular Hamiltonian. The vacuum state (\ref{f1}) is thermal with respect to this notion of time translations.

Naively, we can think the state (\ref{f1}) is locally a Gibbs thermal state with a local inverse temperature given by the coefficient $\beta_x=T_x^{-1}=2\pi x^1$ of the energy density operator in the exponent. This "local temperature"  is completely produced by entanglement with the complementary region $x^1<0$, it is point dependent, and carefully tuned such as to keep all expectation values of operators in $x^1>0$ to coincide with vacuum expectation values. Hence, the thermal interpretation has its limitations, in particular the typical thermal wavelength $\beta_x$ is of the order of the distance from the point to the boundary, and it has the same size as the typical distance in which the temperature changes appreciably, $d\beta/dx=2\pi$. This is necessary in order for local operators not to become really thermalized. However, as we will explain in the next section, there is a precise way to interpret the coefficient of the energy density $T_{00}$ in the modular Hamiltonian as a local inverse temperature using relative entropy. This temperature essentially measures the distinguishability of vacuum from a local high energy excitation. A related interpretation of Unruh temperature has been discussed in \cite{carnot} in terms of a local Carnot temperature.

Another interesting point about (\ref{f1}) is that it holds both for massless and massive fields. In the massive case entanglement is supposed to decay exponentially with the distance to the boundary. However, no such exponential behaviour is seen in the modular Hamiltonian. The reason is that for massive fields a linear increase in $\beta=T^{-1}$ corresponds to an exponential decrease in entropy. This is certainly related to how a local temperature can encode spatial entanglement efficiently in a universal way.

Modular Hamiltonians for other regions have been of recent interest in relation with entropy inequalities and the first law of entanglement \cite{first-law}. This later gives the first order variation of entanglement entropy under variations of the state as the change on the expectation value of the modular Hamiltonian. In the holographic context, the first law of entanglement entropy in QFT has been related to Einstein equations of the dual  gravity theory \cite{einstein}. Modular Hamiltonians have also been recently studied in connection with general properties of CFT \cite{gene}.

It is a natural question whether these features of the Rindler modular Hamiltonian can be generalized to other regions and QFT.  In this work we propose a path to this generalization. We are interested specifically in whether local temperatures  and the related local terms in $K$ can be defined, and when these local terms are proportional to the stress tensor.

We use essentially properties of the relative entropy, and causal evolution. We show that for any region and QFT there is at least a maximum and a minimum local "null temperatures", depending on the point and a null direction. However, for generic Cauchy surfaces and regions different from spheres, the corresponding local contributions to $K$ cannot be always produced by the stress tensor in $d\ge 3$.  We determine the possible structures of the local terms for a free scalar field.

Local terms describe the high energy tail of the reduced density matrix around a point, but are determined by infrared data, such as the geometry of the region. A natural conjecture is that there is a high degree of universality for the null temperatures corresponding to the vacuum state across different QFT, extending the universality of the Rindler case. For example, one could wonder whether the null temperatures depend only on the geometry of the region, and when the maximum and minimum null temperatures actually coincide. We show this last statement is correct for free scalars.

 For free massive scalar and fermion fields in $d=2$ we provide conclusive analytic and numerical evidence for this universality. We show the local term for any multi-interval region is proportional to the stress tensor in this case, with a universal coefficient that is the same for fermions or scalars, and is independent of mass. Our specific arguments for free fields extend to higher dimensions in that the local term should not depend on the mass parameter.

\section{Local temperatures for general regions}

We first give a definition of local temperature appropriate to the Rindler density matrix (\ref{f1}) that will be suitable to be generalized to other regions. As we have recalled the coefficient $\beta=2 \pi x$ of $T_{00}(x)$ in (\ref{f1}) gives the inverse Unruh temperature of an accelerated observer with acceleration $x^{-1}$ that passes through the point $x$. There is no consistent interpretation in terms of locally thermalized expectation values since expectation values of operator in the region coincide with vacuum ones. Hence, we adopt the following strategy involving the relative entropy to define a local temperature.

 The relative entropy between two states reduced to a region $A$ writes
\be
S(\rho^1_A|\rho^0_A)=\Delta \langle K_A \rangle -\Delta S_A
\ee
where $\Delta \langle K_A \rangle=\langle K_A \rangle_1-\langle K_A \rangle_0$ is the variation in the expectation value of the modular Hamiltonian $K_A=-\log \rho^0_A$ of the state $\rho^0$ reduced to $A$, and $\Delta S_A$ is the difference of entropies $S_A^1-S_A^0$.  Hence, if we obtain the state $\rho_1$ by acting on $\rho_0$ with a unitary operator $U_A$ localized in $A$, the variation of entropies will vanish and we have
\be
S(\rho^1_A|\rho^0_A)=\Delta \langle K_A \rangle\ge 0\,.
\ee

\begin{figure}[t]
\centering
\leavevmode
\epsfysize=6cm
\bigskip
\epsfbox{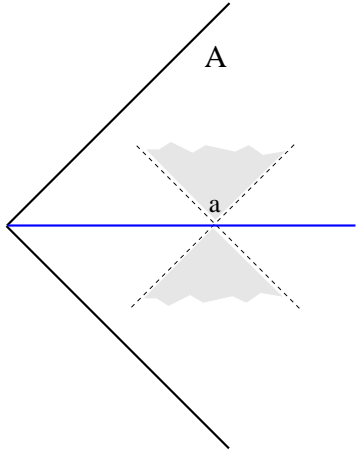}
\caption{We are testing the modular Hamiltonian of a region (here the Rindler wedge $A$) by unitary operator well localized around a point $a$ on Cauchy surface (horizontal line in the figure). The state produced from this unitary acting on the vacuum spreads out in the past and future of the Cauchy surface.}
\label{fig0}
\end{figure}

If we perturb the vacuum with a unitary operator localized in a small region around a point $a$  in  the spatial surface  $x^0=0$ inside the wedge (see figure \ref{fig0}), the expectation value of $T_{00}$ will be different from zero only in this small region. Hence
\be
S(\rho^1_A|\rho^0_A)=\Delta \langle K_A \rangle= 2 \pi \int d^{d-1}x\, x^1\, \langle T_{00}(x)\rangle \sim 2 \pi a^1 E\,,\label{compared}
\ee
where $E$ is the expectation value of the total energy of the excitation. Even if the excitation is not produced by a unitary operator we expect the same result for high energy localized excitations where the change in the entanglement entropy is small compared with (\ref{compared}). However, (\ref{compared}) also holds in the Rindler wedge for localized excitations of arbitrarily small energy expectation value, produced by local unitaries close to the identity.

Formula (\ref{compared}) is the same we would have obtained for the relative entropy of an excitation of energy $E$ above a thermal state of inverse temperature $\beta= 2 \pi a^1$ and the thermal state itself, where these two states are now taken in the full space not restricted to $A$.  This follows because the modular Hamiltonian of the thermal state is $K=\beta H$, proportional to the Hamiltonian $H$.

Relative entropy measures how difficult is to distinguish two states in an operational way. The probability of confounding two states after $N$ judiciously chosen measurements falls to zero as $p\sim e^{-S(\rho^1|\rho^0)N}$ \cite{relative}.
We can say that in trying to distinguish the vacuum and the excitation doing measurements restricted to operators localised in $A$ we find the excitation in the vacuum fluctuations in $A$ with the same probability as in a thermal state with this temperature. Notably, this is independent of the "composition" of the excitation, and only depends on energy, and that is why it is determined by a temperature.

In a QFT the algebras of operators and their states are attached to the domain of dependence of spatial surfaces. That is, the modular Hamiltonian $K$ and the relative entropy between two states, will be the same for any two spatial surfaces with the same causal completion or causal development.
For wedges with general positions (obtained by rotating and boosting the Rindler wedge) the modular Hamiltonian is given by
\bea
K&=&2\pi \int_\Sigma  d\sigma\, \eta^\mu T_{\mu\nu} \xi^\nu\,,\\
\xi^\nu&=&\omega^{\nu\delta} x_\delta \,,
\eea
where $\omega^{\nu\delta}=j^\nu t^\delta-j^\delta t^\nu$, with $j^\nu$, $t^\delta$ two unit spatial and temporal vectors orthogonal to the edge of the wedge. The integral is over any Cauchy surface $\Sigma$ for the wedge and $\eta^\nu$ is the future pointing unit normal to $\Sigma$.
$K$ written in any Cauchy surface is the same operator because it is  the flux of the conserved current $\xi^\mu T_{\mu\nu}$. For a perturbation localized in a small region near the space-time point $a$ inside the wedge we can take $\xi$ as approximately constant in the region where $\langle T^{\mu\nu}\rangle$ is non zero on $\Sigma$, and the local integral $\int_\Sigma d\sigma\, \eta_\nu \langle T^{\mu\nu}\rangle=P^\mu$, the total momentum of the excitation. Then
\be
\Delta K=P_\mu\xi^\mu(a)\,.
\ee
The relativistic generalization of an inverse temperature is given by a vector $\xi^\mu$ such that the Gibbs state writes $\rho\sim e^{-\xi^\mu P_\mu}$. Hence, in this specific sense given by relative entropy,  $\xi^\mu(a)$ has the interpretation of a local inverse temperature vector.

\subsection{Relative entropy inequalities and local temperatures}
Now we try to generalize this structure of local temperatures for other regions. We use relative entropy inequalities.  The relative entropy is always positive and increasing with the region size.
If we excite the vacuum with a local unitary inside $A$, we have for a region $B$ bigger than $A$
\be
\Delta \langle K_B \rangle\ge\Delta \langle K_A \rangle\ge 0\,.
\ee

Let us exemplify these inequalities with CFT and double cones $D(p,q)$ formed by the intersection of the past of a point $q$ with the future of $p$, where $p$ is in the past of $q$.  For the vacuum state of a CFT the modular Hamiltonian in this case is explicitly known and local \cite{conformal}. As in the case of the Rindler wedge it is given by the flux of a conserved current $J^\mu=T^{\mu\nu} \xi_\nu$ on any Cauchy $\Sigma$ surface for $D(p,q)$,
\be
K=\int_\Sigma d\sigma\, \eta^\mu T_{\mu\nu} \xi^\nu\,,\label{fort}
\ee
where $\eta^\mu$ is the future-pointing unit vector normal to the surface and
\be
\xi^\mu(x)=\frac{2\pi}{|q-p|^2}\left((q-x)^\mu (x-p)\cdot (q-p)+(x-p)^\mu (q-x)\cdot(q-p)-(q-p)^\mu (x-p)\cdot(q-x)\right)\,.\label{current}
\ee
$J^\mu$ is a conformal current that vanish on $p$, $q$, and on the spatial boundary of the double cone and $\xi^\mu$ is the associated conformal Killing vector.

 If we perturb the vacuum with a unitary operator localized in the vicinity of a point $a$ inside the double cone, it must be that the expectation value of $K$ is positive in the new state.  For a perturbation localized in a small region near $a$ we can take $\xi$ as approximately constant in the region where $\langle T^{\mu\nu}\rangle$ is non zero on $\Sigma$, and the local integral is again $\int_\Sigma d\sigma\, \eta_\nu \langle T^{\mu\nu}\rangle=P^\mu$. Then
\be
\Delta K=P_\mu\xi^\mu(a)\,.\label{labee}
\ee
  The inequality $\Delta K\ge 0$ implies  that $\xi^\mu(a)$ is a future directed vector for any $a$ inside $D(p,q)$. This can be explicitly checked from (\ref{current}). In the same way, comparing vacuum and the perturbed state, we have from monotonicity of relative entropy that
\be
\xi_{p\prime q\prime}^\mu(a)-\xi_{p q}^\mu(a)
\ee
is also future pointing for any $p^\prime$ to the past of $p$ and any $q^\prime$ to the future of $q$.
 That is, the vector $\xi_{pq}(a)$ for fixed $a$ is increasing in the sense of the causal order determined by the cone of future pointing vectors, for regions increasing under inclusion. This can also be easily checked from (\ref{current}) by perturbing the end points.

The physical interpretation of this fact is very natural. Thinking $\xi(a)$ as a vector determining the inverse local temperature, this increases with larger regions, corresponding to a decrease of temperature as we move the boundaries of the region further away. Entanglement producing this local temperature with more distant degrees of freedom is weaker in vacuum.

Another example where the modular Hamiltonian is known exactly is a massless Dirac field in $d=2$ for a region formed by $n$ disjoint intervals \cite{fermion,nuevo}. This Hamiltonian contains local and non-local terms. The local part writes in null coordinates $x^\pm=x^0\pm x^1$
\be
{\cal K}_{\textrm{loc}}=2\pi\int_{A^+} dx^+\,  f_+(x^+) T_{++}(x^+)+2\pi\int_{A^-} dx^-\,  f_-(x^-) T_{--}(x^-)\,,\label{13}
\ee
where
\be
f_\pm(x^\pm)=\left(\sum_{i=1}^n \frac{1}{x^\pm-a^\pm_i}+\sum_{i=1}^n\frac{1}{b^\pm_i-x^\pm}\right)^{-1}\,,\label{fermion}
\ee
and $(a^\pm_1,b^\pm_1)$, $(a^\pm_2,b^\pm_2)\hdots(a^\pm_k,b^\pm_k)$ are the null coordinates of the end points of the $k$ intervals, written in increasing order. We will encounter these expressions again below arising from a more general argument about QFT theories in $d=2$.

The stress tensor components $T_{++}$ and $T_{--}$ for a localized excitation are positive once integrated over the excitation region and then we should have that $f(x)$ must be positive and increasing with the region size. $f(x)$ is explicitly positive. It is increasing with size because (dropping the $\pm$ for convenience)
\be
\frac{d f(x)}{d b_i}=\frac{f(x)^2}{(b_i-x)^2}>0\,,\hspace{1cm}
\frac{d f(x)}{d a_i}=-\frac{f(x)^2}{(x-a_i)^2}<0\,.
\ee
In comparing regions with different number of components the inequality follows from these ones and the fact that the function $f(x)$ for $k$ intervals tends (for fixed $x$) to the one of $k-1$ components when $b_i-a_i\rightarrow 0$ (and $x\notin (a_i,b_i)$) or when $b_i\rightarrow a_{i+1}$ (in this last case two intervals coalesce to one). Interestingly, in this example, the local temperature $(2\pi f(x))^{-1}$, is the sum of the temperatures (taken with positive sign), $(2\pi)^{-1}(x-a_i)^{-1}$ or $(2\pi)^{-1}(b_i-x)^{-1}$, that would be generated on $x$ by the different boundaries independently, taking as the temperature generated by each boundary independently the one corresponding to the Rindler case.

\bigskip

The monotonicity inequalities have an interesting consequence. Let us take a CFT and any bounded causally complete region\footnote{A causally complete region is the domain of dependence of a spatial surface.} $A$ (not necessarily a double cone) and a point $a$ inside $A$. Consider two double cones $D^+$ and $D^-$ that include $a$, and such that $D^-\subseteq A \subseteq D^+$. Then, for any unitary and well localized perturbation of vacuum around $a$ with momentum $P$ we have
\be
P\cdot\xi_{D^-}(a)\le \Delta \langle K_A \rangle\le P\cdot\xi_{D^+}(a)\,.\label{ffg}
\ee
This highly constraints the contributions of $\Delta K$.\footnote{The information these inequalities provide is insensitive to possible ambiguities in the modular Hamiltonian given by operators seated at the region boundary. These terms do not affect $\Delta K$ produced by a unitary inside the region.}
For any bounded region in any CFT we have that the change of the modular Hamiltonian for an arbitrary localized unitary excitation is positive and bounded above and below by a quantity that depends only on the geometry, and is linear in the excitation momentum. In particular, the relative entropy  with the vacuum state (which measures distinguishability between the two states) cannot exceed a bound proportional to energy.

\begin{figure}[t]
\centering
\leavevmode
\epsfysize=6cm
\bigskip
\epsfbox{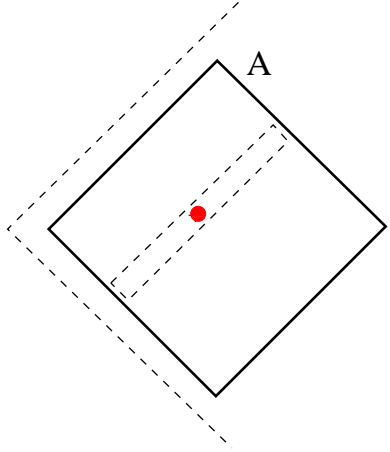}
\caption{The causal region $A$ (black) is included in the Rindler wedge and includes a double cone elongated along a null line (shown with dashed lines). The contributions to
$\Delta K$ for these three regions from a unitary excitation localized near a point (marked with a red circle) are ordered according to the same relations.}
\label{fig1}
\end{figure}

 Note we are testing with relative entropy the local tail of the density matrix around a point, but remarkably the form of this tail depends on infrared details, i.e. the geometry of the boundary far away. The local temperatures, if they persist at all, might change for other states, for example a global thermal state.

For non conformal QFT upper bounds can be derived using the modular Hamiltonian of the Rindler wedge, that has again a universal form.
 To get a lower bound for non conformal theories we can think in a very small double cone inside the region, in such a way the conformal modular Hamiltonian of the UV fix point applies for localized excitations. This double cone can be highly boosted, extending along a null segment inside $A$, and the relevant $\xi$ vector takes a finite value as we take the null limit (the limit of vanishing invariant size of the double cone); see figure \ref{fig1}. In the limit $\xi$ is a null vector. We have for the local excitation at $a$ (in the limit of vanishing size)\footnote{The limit we are considering here is different to the one in \cite{bousso}. There the modular Hamiltonian for a region that converges to a null surface was analysed by taking the limit of a null surface keeping the state constant, while here the perturbation is always concentrated inside the region.}
\be
\Delta \langle K_A \rangle\ge  \pi \frac{(\lambda(a)-\lambda^-)(\lambda^+-\lambda(a))}{(\lambda^+-\lambda^-)}  P_\mu \zeta^\mu \,,\label{lb}
\ee
where $\lambda$ is an affine parameter for the null segment $x^\mu(\lambda)$ passing through $a$, $\zeta^\mu=dx^\mu(\lambda)/d\lambda$ is the corresponding future-pointing null vector, and $\lambda^+$, $\lambda^-$ are the parameters of the extreme points of the segment. The null segment has to be fully included in causal region corresponding to $A$.\footnote{It is not difficult to see that the set of geometric lower bounds coming from (\ref{lb}) for generic theories is not improved thinking in conformal theories where we can use spheres that are not small. Then these lower bounds are completely universal.}

This leads to the interesting observation that if we think in any region $A$ and the limit of localized excitations around $a\in A$, we have
\be
\beta_+^g(a,A,\hat{P},v) E\ge \beta_+(a,A,\hat{P},v) E \ge \Delta K\ge \beta_-(a,A,\hat{P},v) E \ge \beta_-^g(a,A,\hat{P},v) E\,,
\ee
where we have written the inverse "temperatures" $\beta_\pm(a,A,\hat{P},v)$, depending on the point, the causal region,  the momentum direction $\hat{P}$ and velocity $v=|\vec{P}|/E$, as the maximal and minimal values of $\Delta K/E$ attainable for different local excitation composition and energy with the same momentum direction $\hat{P}$ and velocity, and where
$\beta_\pm^g(a,A,\hat{P},v)$ are universal geometrical bounds determined by the geometry alone as described above.\footnote{$\beta_+^g(a,A,\hat{P},v)$ is the minimum of the quantity $\xi_0+\hat{P}^i \xi_i v$ among all $\xi^\mu$ available for upper bounds and $\beta_+^g(a,A,\hat{P},v)$ is the maximum of this quantity for all $\xi^\mu$ available for lower bound.} In general there is a range in $\beta$ allowed by the geometry, and this range depends on the momentum direction and velocity. For the special cases of the Rindler wedge or spheres in CFT this range collapses to a single value for $\beta$ and further, the value of $\beta$ for each momentum direction and velocity precisely comes from the projection of a vector,  eq. (\ref{labee}),
\be
\beta_\pm(a,A,\hat{P},v) E= \xi^\mu(a) P_\mu\,.\label{vect}
\ee
We will see below that both these features are lost for more general regions in $d\ge 3$.

In general we will be interested in local excitations in the limit of large energy because only in this limit we expect  the local temperatures are connected with local features of the modular Hamiltonian. In particular a standard way to produce these high energy excitations is by doing a large boost of a given excitation, and this excitation will be in a null direction, $v\rightarrow 1$. We will then be interested particularly in local inverse temperatures in null directions, that we call simply $\beta_\pm(a,A,\hat{P})$.

If we can produce large energy excitations of nearly null momentum $P_1$ and $P_2$ localized in arbitrarily small regions around $a$ we expect that we can also combine them to be approximately decoupled when they are localized in much smaller sizes that the separation distance.  For a CFT we can produce these states by scaling far away excitations. In that case one could obtain the contribution
\be
\Delta K\simeq \beta(a,A,\hat{P}_1) E_1+\beta(a,A,\hat{P}_2) E_2=\beta(a,A,\hat{Q},v) (E_1+E_2)\,,
\ee
where
\be
\hat{Q}= \frac{E_1 \hat{P}_1+E_2 \hat{P}_2}{|E_1 \hat{P}_1+E_2 \hat{P}_2|}\,,\hspace{.7cm}
v= \frac{|E_1 \hat{P}_1+E_2 \hat{P}_2|}{E_1+E_2} \,.
\ee
This introduces a convexity relation for the possible values of inverse temperatures
\be
p_1 \beta(a,A,\hat{P}_1)+ p_2 \beta(a,A,\hat{P}_2)= \beta(a,A,\hat{Q},v)\label{convex}
\ee
where the two probabilities $p_1$, $p_2$ are such that $p_1+p_2=1$. Analogously one can consider decompositions in several null momenta. This convexity, coming from the one of the momenta in the future cone, carries null momenta to non null momentum $Q=P_1+P_2$.   The relation (\ref{convex}) holds automatically if the inverse temperatures for all directions and velocities come from a vector as in (\ref{vect}). However, if the inverse temperatures in the null directions $\beta(a,A,\hat{P})$ do not come from projections of a vector, (\ref{convex}) leads to a range of possible $\beta(a,A,\hat{Q},v)$ for each non null momentum, even if the null $\beta(a,A,\hat{P})$ are unique (the maximum and minimum coincide). The special features or "composition" of the excitation that in this case leads to a range in $\beta$ is the presence of different clusters moving with different directions. The reason for this range of $\beta$ is that in the future cone the null vectors are "pure" and cannot be decomposed as sums of other vectors while the time-like momenta can be decomposed as sums of null momenta, and this decomposition is highly non-unique. A non null vector in a certain small volume of the future cone can still be decomposed uniquely as a sum of just $d$ fixed null vectors. If we give arbitrary null temperatures for just $d$ null directions we can construct a unique vector $\xi$ which produces these null temperatures and uniquely extend the null temperatures to other null and non null directions. However, if we choose an arbitrary null temperature for an extra $(d+1)^{\textrm{th}}$ null vector, the results for non null vectors will depend on the decomposition.

\subsection{Causal propagation and local temperatures in null directions}
\label{causalprop}

\begin{figure}
\centering
\leavevmode
\epsfysize=6cm
\bigskip
\epsfbox{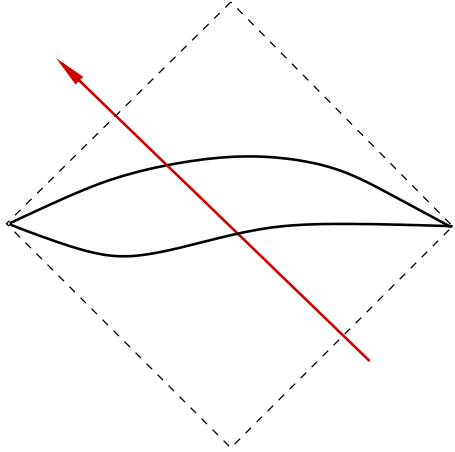}
\caption{The modular Hamiltonian written in two different Cauchy surfaces must have the same expectation value when vacuum is perturbed by an energetic ray which follows an approximately null trajectory.}
\label{fig3}
\end{figure}

Given the high degree of universality of the contribution for $\Delta K$ due to local excitations a natural idea is that a generalization of the Rindler result applies for any region, that is, the local contributions are of the form
\be
\Delta K=P_\mu\xi^\mu(x)\,,\label{labee1}
\ee
for some vector field $\xi^\mu(x)$.
For high energy excitations, this contribution might come from a local term in the modular Hamiltonian analogous to the one in the Rindler wedge,
\be
K_{\textrm{local}}=\int_\Sigma d\sigma\, \eta^\mu T_{\mu\nu} \xi^\nu \,.
\label{local1}
\ee
 This now need not be exactly conserved, it must only give the dominant expectation value for highly localized energetic excitations.

We will see that (\ref{labee1}) is not possible in general, considering only high energy excitations.

Let us take two different Cauchy surfaces $\Sigma_1$ and $\Sigma_2$ for the same region. The full modular Hamiltonian written in these surfaces has to be the same operator giving the same value of $\Delta K$.
 Let us take a state that contains a very energetic excitation with a trajectory sharply localized along a null ray of tangent vector $\chi$ (see figure \ref{fig3}). This type of excitations always exist, and can be constructed by taking a state created by an operator localized in a very small ball and then boosting it.  The momentum of the excitation will be approximately parallel to the null ray, $P^\mu\sim \chi^\mu$.
The expectation value of $K$ computed in the two Cauchy surfaces will be (\ref{labee1}) with $\xi$ written in the vicinity of two points $x_1$, $x_2$ respectively, which are connected by the null ray
\be
x_1-x_2=y \,\chi\,,\label{oltra}
\ee
for some number $y$. The contributions to the modular Hamiltonian are then $P \cdot\xi(x_1)$ and $P \cdot\xi(x_2)$. From the equality of these contributions we get
\be
\chi \cdot(\xi(x_1)-\xi(x_2))=0\,.
\ee
For infinitesimal displacements of the Cauchy surface, this equation, together with (\ref{oltra}), give
\be
\chi^\nu \chi^\mu (\partial_\mu \xi_\nu) =0\,
\ee
for any null vector $\chi$. Then the symmetrized gradient of $\xi$ must be proportional to the metric, and we must have
\be
\partial_\mu \xi_\nu+\partial_\nu \xi_\mu=\frac{2}{d} g_{\mu\nu} (\partial\cdot \xi)\,.\label{edr}
\ee
This is exactly the equation satisfied by infinitesimal coordinate transformations $x^{\mu\,\prime}=x^\mu+\epsilon \,\xi^\mu$ which are conformal transformations of Minkowski metric.

Therefore, eq. (\ref{local1}) must have the same form as the generators of conformal transformations would have if the theory would be conformal, which is not surprising given that the argument only involved the geometry of null rays. However, the unitary modular flow induced by the modular Hamiltonian $e^{- i K \tau}$ must preserve the algebra of operators in the region. This is only possible if the modular flow keeps the operators localized on the spatial boundary of the region fixed. Then it must be that $\xi$ vanishes on $\partial V$. A $\xi$ satisfying (\ref{edr}) can only vanish in spheres for $d\ge 3$ (or the limit case of a plane). As a consequence the local term can be of the form (\ref{local1}) only if the region is a sphere. We know this is the case of a CFT, but in principle this could also be the case of non CFT for spheres in $d\ge 3$.

For $d=2$ the situation is different. Using null coordinates $x^\pm=x^0\pm x^1$ eq.  (\ref{edr}) only requires that $\partial_-\xi_+=\partial_+ \xi_-=0$. Therefore, any functions $\xi_+(x^+)$, $\xi_-(x^-)$ will satisfy the requirement. Then it is possible that the local temperatures for high energy null excitations are produced by a term in the modular Hamiltonian that is just an integral of the stress tensor, for any theory and any number of intervals. We have seen this happens for free fermions, eq. (\ref{13}).

We conclude that we expect that local temperatures in the null directions, that is, in the limit $E\rightarrow \infty$, $P^2$ fixed, to be conserved by propagation, and thus a function of the null ray. We can write for this null directions
\be
 \beta(a,A,\hat{P})=\beta(a^\prime,A,\hat{P})\,,\hspace{.7cm}\textrm{for}\,\, (a-a^\prime)=\lambda (1,\hat{P})\,,
\ee
for some number $\lambda$.
Except in $d=2$ (and spheres) the local temperatures cannot depend on the momentum direction in a simple vectorial form as in (\ref{vect}) for generic regions and Cauchy surfaces. This holds for the maximal and minimal temperatures along null directions, which have to be conserved along the direction of the null ray. This also prevents a simple form (\ref{local1}) for the operator giving place to these temperatures. The reason for this impossibility is that a vector at each point has little information to accommodate all constraints coming from causal propagation and the boundary of the region. We will see in the next Section what kind of operators can do the job for scalars in $d\ge 3$.

Since the temperatures for null directions do not come generically from a vector, it must be that there is a non zero range of temperatures for non null directions, according to the discussion at the end of the previous section. However, it can still be the case that the  temperatures in the null directions, in the limit of large energy, be unique and the maximum and minimum coincide. In any case, it is clear the temperatures in the null directions have a more fundamental nature and more interesting properties than the ones in generic directions.

\subsection{Argument for the universality of local
 temperatures in $d=2$}
 \label{argument}

In $d=2$ we have two null directions and inverse temperatures, which we can call simply $\beta(x,\pm)$, suppressing the dependence on $A$ (this is not to be confused with $\beta^\pm$, here the two signs correspond to the two null directions). These are simply the null components $\xi^\pm$ of the vector $\xi$ in (\ref{local1}) if there is a local term proportional to the stress tensor that gives the relevant contribution. The null temperatures satisfy local equations of motion  $\partial_{\pm} \beta(x,\mp)=0$ that simply state they are conserved along null rays. In Minkowski space this does not give us much information, and the general solution is given by arbitrary functions $\beta(x^\pm,\pm)$.

However, assuming an analytic continuation of the functions $\beta(x,\pm)$ to imaginary time $t\rightarrow -i t$, writing for the complex coordinates $z=x^1+i x^0$, $\bar{z}=x^1-i x^0$, the equations are transformed to $\partial_{\bar{z}} \beta=0$ and $\partial_{z}\bar{\beta}=0$, with analytic and anti-analytic solutions $\beta(z)$, $\bar{\beta}(\bar{z})$.

Near the boundaries of the region $\beta(z)$ should vanish in the way the modular Hamiltonian of Rindler space does (linearly in the distance to the boundary point). Hence, $\beta(z)^{-1}$ should have simple poles, all with the same residue $(2\pi)^{-1}$, at the left boundary points $a_i$ of the intervals and $-(2\pi)^{-1}$ at the right boundary points $b_i$. These are poles of a geometric origin, and for the vacuum state we do not see what could be the origin of other singularities in the complex plane. In particular at the point of infinity  $\beta$ must diverge (giving zero local temperature because it is far away from the boundaries) and $\beta^{-1}$ should go to zero. Then, assuming $\beta^{-1}$ to be analytic everywhere, including infinity, with only simple poles at the boundary, the unique solution is
\be
\beta(z)^{-1}=\frac{1}{2\pi}\sum_i \left( \frac{1}{z-a_i}+\frac{1}{b_i-z}\right)\,.
\ee
This gives a unique $\beta(z)$ which coincides with the one for the free massless fermion described in (\ref{fermion}). The same happens for the antianalytic part.  This local temperatures are also correct for massless scalars \cite{nuevo}.

As the argument leading to this function is independent of the details of the theory, and in particular involves only null propagation, we think this might be the universal form of the local term in $d=2$. We will check this is the case for massive scalars and massive fermions in the following Sections.

An analogous argument in more dimensions would involve an analytic continuation to imaginary time of the statement that $\beta(x,A,\hat{P})$ does not depend on the direction along the ray parallel to $\hat{P}$, or, in other words, that the null temperatures are a functions of null rays.  Questions about holomorphic functions of null rays have been discussed in the literature in terms of twistor space \cite{twistor}. We hope to return to this interesting point in a future work.

\section{Local terms in modular Hamiltonians}

Let us imagine what is the possible form of the operator producing the leading term in $\Delta K$ for an energetic and well localized excitation around a point $a$. We can think locally the geometric coefficients (such as the vector $\xi(a)$ in the Rindler case) can be taken constant, and the contribution is dominated by terms formed by products of operators near $a$. We only need the form of these operators at small distances from $a$ and to each other.  In principle this part will dominate the contribution. Likewise, to evaluate the leading term of the contributions we can use a counting of dimensions corresponding to the UV fix point of the theory. We expect a leading local contribution of the form
\be
K\simeq \sum_k R^\gamma(a) \int_V d^{d-1}x_1\, \hdots d^{d-1}x_k\, G(x_1,\hdots,x_k)\, \phi^{\Delta_1}(x_1)\hdots \phi^{\Delta_k}(x_k)\,.\label{cloud}
\ee
Here $R^\gamma(a)$ stands for a quantity of dimension $-\gamma$ depending on geometric parameters of the region and eventually on the scales of the theory, that is approximately constant around the point of interest.  $G(x_1,\hdots,x_k)$ stands for an homogeneous function of the coordinates of some degree. The $\Delta_i$ are the scaling dimensions of the different field operators (which might contain time derivatives of the field operators) that serve as a generating basis for the operator content of the theory. The integration is on a small region $V$, $a\in V$, on the Cauchy surface.\footnote{A proper mathematical definition of this expression might need some small smearing in the time direction.} Since we can test with excitations increasingly localized in neighbourhoods around any point in this patch, the function $G(x_1,\hdots,x_k)$ has to be homogeneous under scaling around any point. Then it must also be translational invariant, and a function of coordinate differences. Even if we are looking at very localized contributions we cannot use OPE to simplify this expression since the excitation is always assumed to be small inside the cloud of operators (\ref{cloud}).

$K$ is dimensionless, and then the scaling dimension of the integral in (\ref{cloud}) is $\gamma$. This means that taking states scaled down with dilatations by a factor $\lambda$ the expectation value of the integral will scale by a factor $\lambda^\gamma$. For the Rindler wedge or the sphere in a CFT this dimensions is $1$, being proportional to the momentum. Hence we need that $\gamma \le 1$ in order to respect the relative entropy bound. In particular there must be one term such that $\gamma=1$, and this is the leading term we are considering. We would like to constrain the scaling dimensions and the number of operators appearing in (\ref{cloud}) to have a restricted class of possibilities. In principle a large operator dimension can be compensated by large powers on the coordinates in the function $G$.

There is a problem in obtaining additional information from (\ref{cloud}). This is the fact that in principle the sum is over an infinite series, and we do not have any small parameter that select some specific term from the others. Hence it is difficult to understand the meaning of (\ref{cloud}).

 Let us however assume naively that we can analyse each term separately.  If we test with a very localized excitation there will be a contribution to $\Delta K$ where only one of the operators acquire a large expectation value. If this excitation is scaled, the energy will scale as $l^{-1}$, with $l$ the size of the excitation. On the other hand the expectation value $\langle \phi^{\Delta_i}(x)\rangle\sim l^{-\Delta_i}$. Then, there is a contribution to $\Delta K\sim l^{-\Delta_i} l^{d-1}=l^{d-1-\Delta_i}$, where the expectation value of the other operators and the values of $G(x_1,\hdots,x_k)$ are frozen. We then obtain $\Delta_i\le d$; the operators have to be relevant. We can rephrase this constraint as that the cloud of operators scale globally as the energy, but if $\Delta_i>d$, testing with enough detail we could see hard elements inside the cloud, and this is not allowed. In principle there is no problem with this constraint in $d=2$, since the stress tensor can always be producing the local term. For $d>2$ we have seen the stress tensor alone cannot account for the condition of causal propagation of energetic null signals and other relevant operators are needed. This would imply the existence of relevant operators for all CFT in $d>2$. However, the analysis seems to be too naive to arrive to this conclusion, as the following argument shows.

We can take a further step and consider a generic non local term in the modular Hamiltonian, but not necessarily a term that gives the local contributions we were looking for. That is, this term is of the form (\ref{cloud}), but without further conditions on the functions $G$. Non local terms ($k>1$) are needed generically in the modular Hamiltonian. For example, regions with more than one connected component must have non local terms to account for non zero mutual information between the regions, otherwise the density matrix factorizes.
Suppose we test with a state formed by $k$ localized excitations at the same time, located at fixed distances between them, and scale the energies of these excitations independently as $l^{-\alpha_1},\hdots,l^{-\alpha_k}$. We get from monotonicity of relative entropy that
\be
\prod_{i=1}^k l^{-\alpha_i (\Delta_i-(d-1))} \le \max(l^{-\alpha_i})\,.
\ee
From this, in the limit of small $l$, we get
\be
\sum \alpha_i \Delta_i \le  (d-1) \sum \alpha_i + \max \alpha_i  \hspace{.6cm} \forall \alpha_i> 0\,.
\ee
Choosing all $\alpha_i$ equal we have in particular
\be
\sum_i \Delta_i \le  k (d-1)+1\,.\label{bbv}
\ee
This last inequality requires there is at least one $\Delta$ satisfying
\be
\Delta\le (d-1)+k^{-1}<d \,,\label{ultim}
\ee
 which is strictly relevant.\footnote{
For $d=2$ CFT we can make this argument more precise. We have a unitary representation of the infinite dimensional group of reparametrizations of the two null coordinates (see for example \cite{confor}). We can use these unitaries in our tests of relative entropy. We are thinking in writing the modular Hamiltonian in a null surface in terms of chiral fields on this surface. All coordinates in the following will be null, let say $x^+$ coordinates.  Let $x^{+\,\prime}=f(x^+)$ be one such reparametrization which is equal to $x^+$ outside some small interval, and it is invertible, $f^\prime(x^+)\neq 0$. This defines a localized unitary $U_f$ in the $x^+$ axis such that
\be
U_f^\dagger T_{++}(x^+) U_f= (f^\prime(x^+))^2 T_{++}(f(x^+))+\frac{c}{24 \pi} \left(\frac{3}{2}\left(\frac{f^{\prime\prime}}{f^\prime}\right)^2-\frac{f^{\prime\prime\prime}}{f^\prime} \right)\,,
\ee
where $c$ is the central charge and the second term in the right hand side is called the Schwarz derivative. This last term gives the expectation value of the stress tensor in the state $U_f|0\rangle$. Its integral is the total energy and has to be positive,
\be
\frac{c}{24 \pi}\int dx^+\,  \left(\frac{3}{2}\left(\frac{f^{\prime\prime}}{f^\prime}\right)^2-\frac{f^{\prime\prime\prime}}{f^\prime} \right)=\frac{c}{48 \pi}\int dx^+\,  \left(\frac{f^{\prime\prime}}{f^\prime}\right)^2\,.\label{energy}
\ee
On a product of primary fields these unitaries act covariantly as
\be
U_f \phi^{\Delta_1}(x_1)\hdots \phi^{\Delta_k}(x_k)U_f^\dagger =f^\prime(x_1)^{\Delta_1}\hdots f^\prime(x_k)^{\Delta_k} \,\,\phi^{\Delta_1}(f(x_1))\hdots \phi^{\Delta_k}(f(x_k))\,.
\ee
Then, for a generic homogeneous term we have
\be
\Delta K=R(a) \int dx_1\,\hdots dx_k\, G(x_1,...,x_k)\, \left(f^\prime(x_1)^{\Delta_1}\hdots f^\prime(x_k)^{\Delta_k} C_0(f(x_1),\hdots,f(x_k))-C_0(x_1,\hdots,x_k)\right)\,,\label{exx}
\ee
where $C_0(x_1,\hdots,x_k)=\langle 0|\phi^{\Delta_1}(x_1)\hdots \phi^{\Delta_k}(x_k)|0\rangle$.
 We can take a function $f(x)$ that is different from $x$ only in small intervals of fix size $b$ around the points $\bar{x}_1,\hdots,\bar{x}_k$, but where the derivative $f^\prime$ acquire much larger values $l^{-1}$. We can take $f(x)$ to be formed piecewise by exponentials $\sim e^{\pm x/l }$ to make this shape. Exponentials locally minimize the contribution to the energy (\ref{energy}).  The energy will scale with fixed $b$ as
$ E\sim l^{-1}$. The expression (\ref{exx}) have a contribution for $x_i$ around $\bar{x}_i$ where we can take $ G(x_1,...,x_k)\sim G(\bar{x}_1,...,\bar{x}_k) $ and $ C_0(x_1,...,x_k)\sim C_0(\bar{x}_1,...,\bar{x}_k) $ as roughly constant, and $\Delta K$ will essentially scale as
$ \Delta K\sim (l)^{-\sum_1^k \Delta_i} \times l^k$,
giving the same bounds on dimensions as in (\ref{ultim}).
}
This argument seems to suggest that any theory would need to contain relevant operators. However, some $d=2$ models are known which do not contain relevant operators \cite{monster}. This tells of the limitations on the interpretation of the infinite sum (\ref{cloud}).
	
In the particular case of free fields we can go beyond in this analysis because we know the vacuum state is Gaussian and the sum in (\ref{cloud}) contains only quadratic term in the fields. We are going to analyse in more detail the scalar fields in general dimensions and in the next Section turn attention to the effect of a mass for free fields.

\subsection{Structure of local terms for free scalars}

Let us consider a free massive real scalar field and creating a state acting on vacuum with a unitary localized operator. We can use the coherent states
\be
|\alpha\rangle=e^{i\int dx\, \tilde
{\alpha}(x) \phi(x)}|0\rangle\,,
\ee
where $\tilde{\alpha}(x)$ is real. We have
\be
[\phi(x),e^{i\int dx \,\tilde{\alpha}(y) \phi(y)}]=e^{i\int dx\, \tilde{\alpha}(y) \phi(y)} \int dy\,\tilde{\alpha}(y)\, i [\phi(x),\phi(y)]=e^{i\int dx\, \tilde{\alpha}(y) \phi(y)} \alpha(x)\,,\label{conmu}
\ee
where we have written
\be
\alpha(x)=-\int dy\,\Delta(x-y)\tilde{\alpha}(y)\,,\hspace{1cm} \Delta(x-y)=-i [\phi(x),\phi(y)]\,.
\ee
The function $\Delta(x-y)$ is a real antisymmetric solution of the homogeneous equations of motion which vanishes outside the light-cone. Hence $\alpha(x)$ is also a solution of the equations of motion $(\partial^2 +m^2)\alpha(x)=0$, which vanishes outside the future and past of the support of $\tilde{\alpha}(x)$. Our notation $|\alpha\rangle$ for the state is due to the fact that this state depends on the wave $\alpha$ rather than function $\tilde{\alpha}$ used to create it. We then have
\be
e^{-i\int dx\, \tilde
{\alpha}(x) \phi(x)} \phi(x)e^{i\int dx\, \tilde
{\alpha}(x) \phi(x)}=\phi(x)+\alpha(x)\,.\label{hjh}
\ee

Using this we have for the two point function
\be
\langle\alpha |\phi(x)\phi(y)|\alpha\rangle=\langle 0|\phi(x)\phi(y)|0\rangle + \alpha(x)\alpha(y)\,.\label{44}
\ee
The vacuum subtracted two point functions are purely classical, given by the replacement of the operator $\phi(x)$ by the classical function $\alpha(x)$. In particular for the stress tensor
\be
\langle\alpha |T_{\mu\nu}|\alpha\rangle-\langle 0 |T_{\mu\nu}|0\rangle=\partial_\mu \alpha(x) \partial_\nu \alpha(x)-g_{\mu\nu}\frac{1}{2}(\partial_\beta \alpha(x) \partial^\beta \alpha(x)+m^2 \alpha(x)^2 )\,.\label{tmu}
\ee

From (\ref{hjh}) we know that scaling the function $\alpha\rightarrow \lambda \alpha$ without scaling the coordinates, the expectation values of products of $n$ fields will have a term scaling like $\lambda^n$. Since the expectation value of the stress tensor scales as $\lambda^2$ we can deduce from the inequalities for the Rindler wedge that the vacuum modular Hamiltonian for any bounded region has to be quadratic in the fields. Of course, we can also deduce this from the fact that correlation functions for the free field are Gaussian \cite{Peschel}.

This great simplification allow us to write the modular Hamiltonian in any Cauchy surface as a bilinear in the fields and momentum operators. From (\ref{44}) it also follows that if one of the coordinates of the two fields involved in $K$ falls outside of the support of $\alpha$, this non local part does not contribute to $\Delta K$.
The leading local contribution around some small patch $V$ around the point $a$ where we want to test with our coherent state, scaling for high energies in the same way as the Rindler Hamiltonian, must have a general form
\be
 \int_V d^{d-1}x\,d^{d-1}y\, (M(x-y)\phi(x)\phi(y)+N(x-y)\dot{\phi}(x)\dot{\phi}(y)+Q(x-y) \phi(x) \dot{\phi}(y))\,.\label{forma}
\ee
 $\dot{\phi}$ is the derivative of the field in the normal direction to the Cauchy surface, and $x,y$ are local Cartesian coordinates on this surface. The fields $\phi$ and $\dot{\phi}$ form a complete basis for operators. All kernels are real, and $M$ and $N$ are symmetric. Since numerical terms on the modular Hamiltonian are irrelevant in our analysis we do not care about the ordering of operators.
 These kernels depend on the point $a$, though we are not writing this explicitly.
For a surface at $t=0$ in a time reflection symmetric causal region the kernel $Q(x-y)=0$. For the vacuum subtracted contribution due to a coherent state we have to replace $\phi(x)$ by $\alpha(x)$ in (\ref{forma}).

In order that this contribution has the same scaling than the Rindler Hamiltonian as we scale the coordinates in $\alpha(x)\rightarrow \alpha(\lambda x)$ we need the kernels $M$, $N$, and $Q$ to be exactly homogeneous distributions of scaling dimensions $d+1$, $d-1$, and $d$, respectively. A homogeneous distribution $H(x^1,x^2,...,x^k)$ of $k$ real variables and dimension $\Delta$ is such that $H(\lambda x^1,\lambda x^2,...,\lambda x^k)=\lambda^{-\Delta} H(x^1,x^2,...,x^k)$. The distributions $M,N,Q$ are functions of $d-1$ variables. Homogeneous distributions are classified \cite{gelfand}. For a number $d-1$ of variables and dimensions $d-1$, $d$, and $d+1$ they include combinations of derivatives of the delta function (which has dimension $d-1$), and regularized power functions:
\bea
M(x)&=& - \frac{1}{2}a_{ij} \partial_i \partial_j \delta(x)+\frac{1}{2}|x|^{-(d+1)} m(\hat{x})\,, \\
N(x)&=&\frac{b}{2} \,\delta(x) +\frac{1}{2} |x|^{-(d-1)} n(\hat{x}) \,,\\
Q(x)&=&  c_i \partial_i \delta(x)+|x|^{-d}  q(\hat{x})\,.
\eea
$m(\hat{x})$, $n(\hat{x})$ and $q(\hat{x})$ are functions on the unit sphere such that \cite{gelfand}
\be
\int d\Omega \, m(\hat{x}) \hat{x}^i \hat{x}^j  =0\,,\hspace{.5cm}
\int d\Omega \, n(\hat{x})    =0\,, \hspace{0.5cm}
\int d\Omega\, q(\hat{x}) \hat{x}^i  =0\,.\label{a}
\ee
As explained above, all quantities $a_{ij}$, $b$, $c_i$, $m(\hat{x})$, $n(\hat{x})$, $q(\hat{x})$, are slowly varying functions of the point $a$ which are considered as constants in the small patch $V$ of interest, and they all have dimensions of length.

The terms involving the delta functions correspond to local operators in the modular Hamiltonian. An integral of stress tensor components produce terms of this kind. The power functions do not come from integrals of local operators. As argued in the preceding Section for general regions and Cauchy surfaces these terms are needed in dimensions $d\ge 3$. Notice the functions $m$, $n$, $q$, depend on $d-2$ variables, and thus have the potential to encode the details of the boundary surface of the region $A$. Without this freedom we have seen it is not possible to satisfy the constraints of causal propagation and vanishing of the modular Hamiltonian at the boundary $\partial A$.

It is interesting to note that other regularizations of the powers functions with the right dimensions exist, but they are not fully homogeneous. For example, instead of the principal value of $1/x$ we can use $1/|x|$. The action of this distribution on a test function $\alpha(x)$ is defined by
\be
\lim_{\epsilon\rightarrow 0}\left(\int_{|x|>\epsilon} dx\, \frac{\alpha(x)}{|x|}+2 \alpha(0) \log(\epsilon)\right)\,.
\ee
Applying this distribution to the scaled function $\alpha(\lambda x)$ we get instead
\be
\lim_{\epsilon\rightarrow 0}\left(\int_{|x|>\epsilon} dx\, \frac{\alpha(x)}{|x|}+2 \alpha(0) \log(\epsilon)\right)-2 \alpha(0)\log(\lambda)\,,
\ee
which is not invariant under scaling but has a logarithmic correction. In the present context this would lead to a violation of the relative entropy inequalities.\footnote{However, the distribution $1/|x-y|$ does make perfect sense as a correlator of scalar fields of dimension $1/2$ in Euclidean CFT. The reason is that in Euclidean field theory the distributions act over a space of test functions that vanish (with all their derivatives) at the coincidence point $x=y$. On this test function space (that we cannot use here), this distribution is homogeneous under scaling of coordinates as a CFT correlator of primary fields must be.} This is the reason behind the conditions (\ref{a}) in general dimensions.

\bigskip

Now we see what information we can obtain from the propagation between different Cauchy surfaces together with the relative entropy inequalities. We use coherent states again as test states. Taking into account $\alpha$ is a real solution of the Klein-Gordon equation\footnote{These are not particle excitations but coherent states corresponding to the wave $\alpha(x)$ which can have positive or negative energy, while the quantum state is of positive energy as it must be, see eq. (\ref{tmu}).}  we write it as
\be
\alpha(x)=\int \frac{d^{d-1}p}{(2\pi)^{(d-1)/2}\sqrt{2|\vec{p}|}}\,\left(a_{\vec{p}}\,e^{-i |\vec{p}| t+i \vec{p}\cdot \vec{x} }+a_{\vec{p}}^* \,e^{i |\vec{p}| t-i \vec{p}\cdot \vec{x} }  \right)\,.
\ee
On the spatial surface we have
\bea
\alpha(\vec{x})&=&\int \frac{d^{d-1}p}{(2\pi)^{(d-1)/2}\sqrt{2|\vec{p}|}}\,\,\left(e^{i \vec{p}\cdot \vec{x}} a_{\vec{p}}+e^{-i \vec{p}\cdot \vec{x}} a_{\vec{p}}^*\right) \,,\label{fie}\\
\dot{\alpha}(\vec{x})&=&-i \int \frac{d^{d-1}p}{(2\pi)^{(d-1)/2}\sqrt{2|\vec{p}|}} \,|\vec{p}|\,\left(e^{i \vec{p}\cdot \vec{x}} a_{\vec{p}}-e^{-i \vec{p}\cdot \vec{x}} a_{\vec{p}}^*\right)\,.\label{mom}
\eea
  We are using a massless approximation since we are interested in high energy excitations, and set $p^0=\pm |\vec{p}|$.

The Fourier transform of a homogeneous distribution of dimension $\Delta$ is again a homogeneous distribution, of dimension
$n-\Delta$ in the conjugate variables, with $n$ the dimension of the space.  We have for a generic homogeneous distribution
\be
\hat{H}(p)=\int dx^n\, e^{i p x} \frac{h(\hat{x})}{|x|^\Delta}=\frac{\hat{h}(\hat{p})}{|p|^{n-\Delta}}\,.
\ee
The functions $h(\hat{x})$ and $\hat{h}(\hat{p})$ have the same dimension and are non-locally related by
\be
\hat{h}(\hat{p})=\lim_{\epsilon\rightarrow 0^+}\Gamma[n-\Delta]\int d\Omega_x\, (\epsilon-i \hat{p}\cdot\hat{x})^{\Delta-n} h(\hat{x})\,.\label{dhl}
\ee
For $\Delta=n,n+1,n+2$, corresponding to our kernels $n$, $q$, $m$, we have to define these integrals by analytic continuation in $\Delta$, and the limit is well defined because of the conditions (\ref{a}). The relevant kernel of $\hat{p}\cdot\hat{x}$ in (\ref{dhl}) is logarithmic in these cases.

Using (\ref{fie}) and (\ref{mom}) into (\ref{forma}) we have for the local contributions to $\Delta K$ written in momentum space,
\be
\Delta K = \frac{1}{2} \int d^{d-1}p\,\,|\vec{p}|  \left(\begin{array}{c}
a_{\vec{p}}^* \\ a_{-\vec{p}}
\end{array}\right)^T\left(
\begin{array}{cc}
\beta(\hat{p}) & \gamma(\hat{p})\\ \gamma^*(\hat{p}) & \beta(-\hat{p})
\end{array}\right)\left(
\begin{array}{c}
a_{\vec{p}} \\ a_{-\vec{p}}^*
\end{array}
\right) \,,\label{62}
\ee
where we have defined
\be
\beta(\hat{p})=\frac{1}{2}(\tilde{m}(\hat{p})+\tilde{n}(\hat{p})+\tilde{q}(\hat{p})+\tilde{q}(\hat{p})^*)\,,\hspace{.7cm}
\gamma(\hat{p})=\frac{1}{2}(\tilde{m}(\hat{p})-\tilde{n}(\hat{p})+\tilde{q}(\hat{p})-\tilde{q}(\hat{p})^*)\,,
\ee
with
\be
\tilde{m}(\hat{p})=a_{ij} \hat{p}_i \hat{p}_j+ \hat{m}(\hat{p})\,,\hspace{.6cm}\tilde{n}(\hat{p})=b+\hat{n}(\hat{p})\,, \hspace{.6cm} \tilde{q}(\hat{p})=c_i \hat{p}_i+i \hat{q}(\hat{p})   \,.\label{dopl}
\ee
We recall $a_{ij}$, $c_i$, and $b$ are real,
$\hat{m}(\hat{p})$ and $\hat{n}(\hat{p})$ are real and symmetric while $\hat{q}(\hat{p})=\hat{q}(-\hat{p})^*$. This gives   $\tilde{m}(\hat{p})$ and $\tilde{n}(\hat{p})$ real and symmetric while $\tilde{q}(\hat{p})=-\tilde{q}(-\hat{p})^*$.
In consequence $\beta(\pm\hat{p})$ are real and $\gamma(\hat{p})=\gamma(-\hat{p})$ symmetric.
We see from (\ref{dopl}) that the full kernels on the unit sphere when written in momentum space do not have the restrictions (\ref{a}). These restrictions are precisely lifted by the addition of the delta function kernels in coordinate space.

The matrix in (\ref{62}) has to be positive semi-definite for each value of $\hat{p}$ to have positive relative entropy.\footnote{The contributions to the integral for $\vec{p}$ and $-\vec{p}$ are identical and independent of the contributions in other directions.} That is, $\beta(\pm\hat{p})\ge 0$, and $\beta(\hat{p})\beta(-\hat{p})\ge|\gamma(\hat{p})|^2$.

Now, the projection of the momentum of the excitation in some future-directed time-like vector $\xi$ writes
\be
\xi_\mu P^\mu= \frac{1}{2}\int d^{d-1}p\,\,|\vec{p}|  \left(\begin{array}{c}
a_{\vec{p}}^* \\ a_{-\vec{p}}
\end{array}\right)^T\left(
\begin{array}{cc}
\xi^0 + \xi^i \hat{p}_i & 0\\ 0 & \xi^0-\xi^i \hat{p}_i
\end{array}\right)\left(
\begin{array}{c}
a_{\vec{p}} \\ a_{-\vec{p}}^*
\end{array}
\right) \,. \label{fth}
\ee
 This quantity constitutes an upper bound and a lower bound on (\ref{62}) for different $\xi$.  The comparison between (\ref{fth}) and (\ref{62}) can be done in each independent direction of the momentum, implying that the difference of matrices have to be positive definite.

For an excitation highly concentrated in a single momentum direction $\hat{p}$, $a_{\vec{p}}\neq 0$, $a_{-\vec{p}}\sim 0$, we recognise from the comparison between (\ref{fth}) and (\ref{62}) that $\beta(a,\hat{p})$ (where we explicitly recall the dependence on the point $a$) is exactly what we have called inverse temperature in the null direction determined by $\hat{p}$ at the point $a$. This must be preserved by causal propagation, being the same for any point in the null ray passing through $a$ in direction $\hat{p}$,
\be
\beta(a,\hat{p})=\beta(b,\hat{p})\,, \hspace{.7cm} a-b=\lambda (1,\hat{p})\,.
\ee
If we excite $a_{\vec{p}}$ and $a_{-\vec{p}}$ together, and look at the contribution in another Cauchy surface, the possible presence of a non zero $\gamma(\hat{p})$ points to non local correlations for the contributions of the two wave packet in this new surface. These cross terms still scale as the energy.

We can also write the local term (\ref{forma}) in terms of creation and annihilation operators by introducing the corresponding expansions of the field operators, with the approximation that the kernels are taken with a constant form in the small region of interest. The result is the same as (\ref{62}) where now $a_{\vec{p}}$, $a_{\vec{p}}^*$ are annihilation and creation operators, and the ordering is not relevant since we are not interested in constant terms in the modular Hamiltonian. We see that for excitations of any kind, not necessarily coherent states, with momentum highly concentrated in a null direction, the contribution of the modular Hamiltonian is
\be
K\simeq \beta(a,\hat{p})\int d^{d-1}p\, \, |\vec{p}|  a^*_{\vec{p}} a_{\vec{p}} \,.
\ee
This indicates that there is no range in the null temperatures, and we have that the maximum and minimum coincide,
\be
\beta^+(a,\hat{p})=\beta^-(a,\hat{p})=\beta(a,\hat{p})\,.
\ee

\bigskip

For $d=2$,  $\hat{m}(\hat{p})$ and $\hat{n}(\hat{p})$ can have two possible values for the two null directions, but they have to be symmetric and because of (\ref{a}) the two values have to sum zero. Hence they vanish. When going back to the coordinate representation for example the contribution of $m$ would be
\be
\int dx\,dy\, \partial_x \phi(x) \frac{1}{x-y} \partial_y \phi(y)\,,\label{ghjk}
\ee
where the distribution is understood in principal value regularization. This is antisymmetric, and then (\ref{ghjk}) vanishes.
For intervals on the $t=0$ line $\tilde{q}(\hat{p})=0$ by time reflection symmetry. Then we have generically a local term in this case
\be
\int dx\, (f_1(x)\dot{\phi}^2 + f_2(x)(\partial_x \phi)^2)\,.\label{saw}
\ee
This is the integral of a local operator but still not necessarily the stress tensor since it may be $f_1\neq f_2$. For the massless limit the chiral components $\partial_{x^\pm}\phi$ decouple, and this is consistent with (\ref{saw}) only for $f_1=f_2$. Then we have a term proportional to $T_{00}$.

 The argument in Section \ref{argument} hinted that only the stress tensor appears, and the coefficient functions should be universal and given by (\ref{13}). This is the case of massless free fields \cite{fermion,nuevo}. But actually we have not proved that these coefficients are independent of mass or that something like $f_1\neq f_2$ cannot happen when there is a mass. We turn to this analysis in more detail in the next Sections.

\section{Local terms for free fields in $d=2$}
In this Section we argue that the leading local terms in the modular Hamiltonian for free fermion and scalar fields in $d=2$ for any number of intervals have the same form given by (\ref{13}), (\ref{fermion}), independently of the field mass. This is in accordance with the argument in Section \ref{argument} for universality of the local terms across different theories based on causal propagation and analyticity on the Euclidean plane. We will further check this result numerically in the next Section.

 A complete description of the reduced density matrix for a free field can be achieved if we can diagonalise the correlator kernel in the region of interest. For example, for a Dirac field on $n$ intervals on the $x^0=0$ surface the modular Hamiltonian is
 \be
K=\int_A dx\,dy\, \psi^\dagger(x) H(x,y) \psi(y)\,,\label{mama}
\ee
where
\be
H=-\log(C^{-1}-1)\,,\label{kernel}
\ee
and
\be
C(x-y)=\langle 0|\psi(x)\psi^\dagger(y)|0\rangle
\ee
is the equal time correlator kernel.

In \cite{fermion} we obtained a complete diagonalisation
of $C$ in the massless case, in a region $A=(a_1,b_1)\cup (a_2,b_2)\cup\hdots \cup (a_n,b_n)$ formed by $n$ intervals on the $x^0=0$ line in $d=2$. A complete set of eigenfunctions for each chirality (see \cite{fermion,nuevo}) is
\be
\psi_s^k(x)=\frac{1}{N_k}\frac{\prod_{i\neq k}(x-a_i)}{\sqrt{-\prod_{i=1}(x-a_i)(x-b_i)}}e^{-i s \omega(x)}\,,\hspace{.6cm}
w(x)=\ln \left( -\prod_{i=1}^n \frac{x-a_i}{x-b_i}\right)  \in (-\infty,\infty)\,,\label{solunorm}
\ee
with the normalization factor
\be
N_k=\sqrt{2\pi}\left(\frac{\prod_{i\neq k}(a_k-a_i)}{\prod_{i}(a_k-b_i)}\right)^{1/2}\,.
\ee
There are $n$ degenerate eigenvectors for $k=1,\hdots,n$, with eigenvalue parametrized by $s\in(-\infty,\infty)$,
\be
\int_A dx\, C(x-y) \psi_s^k(y)= \frac{1}{2}(\tanh(\pi s)+1) \psi_s^k(x)\,.\label{eig}
\ee
 The eigenvectors are normalized according to
\be
\int_A dx \, \psi_s^{k*}(x)\psi_s'^{k'}(x)=\delta_{k,k'}\delta(s-s').\label{normi}
\ee

Therefore, using (\ref{kernel}) the kernel $H$ of the modular Hamiltonian writes
\be
H(x,y)=\int ds\,\sum_{k=1}^n \psi_s^k(x) (2\pi s) \psi_s^k(y)^*\,. \label{asado}
\ee
The result contains local and non local terms. The local term is
\be
H(x,y)_{\textrm{loc}}=2 \pi f(x) \delta^\prime(x-y)+\pi f^\prime(x)\delta(x-y)\,,\label{asa}
\ee
with $f$ given by (\ref{fermion}).
The delta function just makes the delta prime term Hermitian. Once this is plugged into (\ref{mama}) it gives the term proportional to the stress tensor (\ref{13}). See \cite{nuevo} for details. The point we want to make here is that this local term proportional to $\delta^\prime(x-y)$ appears because the dependence on $s$ of the integrand in (\ref{asado}) is of the form $s e^{i(w(x)-w(y))}$, what gives a  $\delta^\prime(w(x)-w(y))$. This singular term will not be modified if the eigenfunctions $\psi^k_s(x)$ change but have the same limit for $s\rightarrow \pm \infty$. Hence, for the massive case we are going to argue that in this limit of large $|s|$ the eigenfunctions of $C$ go to the eigenfunctions of the massless problem. In that case the massive Hamiltonian has the same leading local term as the massless one.

However, analysing the effect of the mass in the eigenvalue problem (\ref{eig}) in the limit of large $s$ is complicated. This is an integral equation, where the mass enters changing the correlator kernel in an important way. This correlator will be written in terms of Bessel functions of $m(x-y)$, where this product cannot be taken small in general. The limit of large $|s|$ corresponds to eigenvalues of $C$ near $0$ or $1$, what does not give any evident clue.

In order to proceed, we rewrite the eigenvalue problem (\ref{eig}) into a different form. The proof of the following statement will be presented in \cite{nuevo}.  Let $S(x)$ be a solution of the massive Euclidean Dirac equation in the plane, smooth everywhere except at a cut located on the region $A$ in the line $x_2=0$, where the function jumps with a fixed factor, according to the following boundary condition
\be
S^{+}(x_1)=\lim_{x_2\rightarrow 0^+} S(x_1,x_2)=-e^{2 \pi s} \lim_{x_2\rightarrow 0^-} S(x_1,x_2)=-e^{2 \pi s}\, S^-(x_1)\,, \hspace{1cm} x_1\in A. \label{disc}
\ee
Then it follows that $S^+(x)$, $x\in A$, is a solution of the eigenvalue problem (\ref{eig}). This transform the problem of integral equations into one of partial differential equations with boundary conditions.

The question is then what happens for the solution $S(x)$ as we make the jump on the cut arbitrarily large, $s\rightarrow \pm \infty$. This large jump factor will necessarily produce large gradients in $S(x)$ in the region of the plane near the cut. Large gradients reduce the effect of the mass in the Dirac equation. This leads to the desired result that the eigenfunctions are independent of mass in the large $|s|$ limit.

The complete massive solution of this problem is not known. But let us exemplify the idea with the simple problem of a cut in the line $x_1>0$. In polar coordinates $(r, \theta)$ the solution of the equation for one of the spinor components that satisfy  $(-\nabla^2+m^2)S(\vec{x})=0$,  with the boundary conditions (\ref{disc}), is proportional to
\be
S= m^{1/2-i s} K_{i s-1/2}(m r)e^{(s+i/2) \theta}\,,
\ee
where $K$ is the Bessel function. Notice $s$ plays the role of an angular momentum. For large angular momentum and a fixed $r$, $|s|\gg m r$, the solution is independent of the mass
\be
S(r,m)\rightarrow \textrm{cons}\,\, r^{-1/2 + i s} e^{(s+i/2) \theta} \hspace{.7cm} |s|\gg m r\,,
\ee
disregarding if $mr$ is big or small.

Another way to convince oneself the mass does not play a role in large $|s|$ eigenvectors is to think that the boundary condition in (\ref{disc}) can be imposed by adding a potential to the Dirac equation localised on the cut,
\be
(\gamma_\mu \partial_\mu +m + \gamma^0 (i\pi+2 \pi s) \Theta_A(x_1) \delta(x_2)) S(\vec{x})=0\,,\label{dirde}
\ee
where $\Theta_A(x_1)$ is equal to $1$ on the cut and $0$ elsewhere.
This potential just imposes the boundary conditions. With large $|s|$ we have a large potential term on $A$, and this again makes the mass to be negligible in the region near the cut.

\begin{figure}[t]
  \begin{subfigure}[b]{0.5\linewidth}
    \centering
    \includegraphics[width=0.75\linewidth]{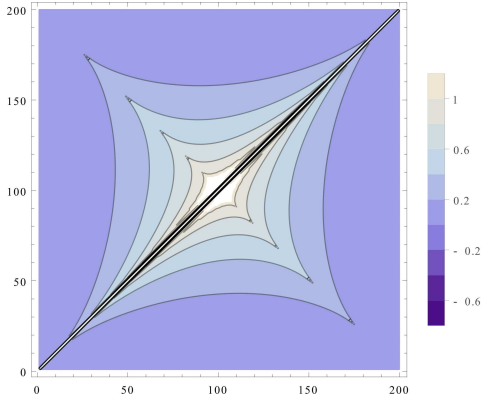}
    \caption{Contour plot of the matrix $N$ for $L=200$ and \newline mass $mL=1$.}
    \label{N1:a}
    \vspace{4ex}
  \end{subfigure}
  \begin{subfigure}[b]{0.5\linewidth}
    \centering
    \includegraphics[width=0.9\linewidth]{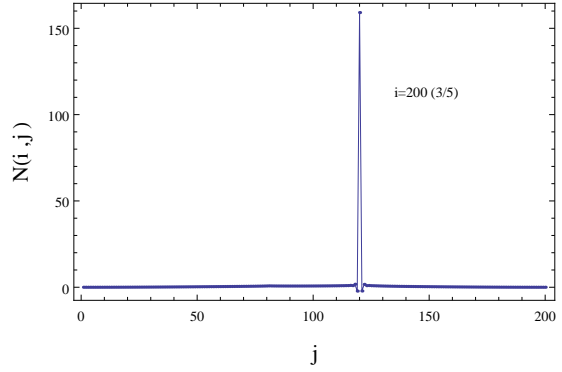}
    \caption{  Delta function: Modular Hamiltonian kernel $N(\frac{3}{5} 200,j)$ for an interval of length $L=200$ and
     mass $mL=1$.}
    \label{N1:b}
    \vspace{4ex}
  \end{subfigure}
  \begin{subfigure}[b]{0.5\linewidth}
    \centering
    \includegraphics[width=0.75\linewidth]{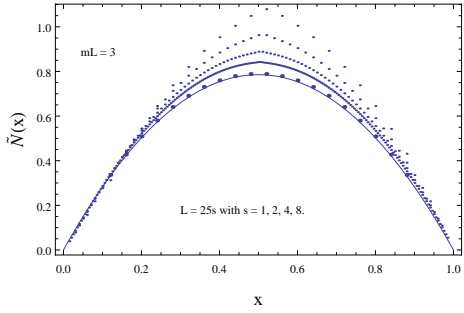}
     \caption{Numerical values of the contributions to the delta
     \newline function $(N(x,x)+N(x,x+1)+N(x,x-1))/L$ (scaled \newline
      to $L=1$),  for $mL=3$ with $L=25 \lambda$ and $\lambda=1,2,4,8$.
 \newline The solid line is the theoretical curve and the points near\newline  it is the fit to the continuum limit (using $\lambda=1,...,8$).}
       \label{N1:c}
  \end{subfigure}
  \begin{subfigure}[b]{0.5\linewidth}
    \centering
      \includegraphics[width=0.75\linewidth]{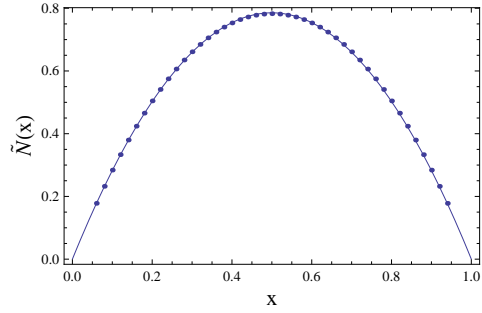}
    \caption{Numerical continuum limit (points) and theoretical (solid line) result  $\pi x(L-x)/L$ with $L=1$. We used $mL=1$.\newline
    \newline\vspace{.4cm}}
   \label{N1:d}
  \end{subfigure}
  \caption{Scalar in a single interval. Kernel $N$ for a scalar field. }
  \label{N1}
\end{figure}

These arguments apply as well for the case of higher dimensions, since eqs. (\ref{eig}) and (\ref{disc}) are still valid, and the boundary values of $S(x)$ give the eigenfunctions of the correlator. In this sense we expect again the eigenfunctions of the massive problem converge to the ones of the massless one for $|s|\gg m R$, for the typical scale $R$ of the region. We also expect that only the large $|s|$ part of the problem is related to local terms, since singular terms in the modular Hamiltonian kernel cannot appear without an improper integral. Hence, we expect that local terms in any dimensions will be insensitive to mass. This would imply for example the leading high energy local term on a sphere is the usual one for a CFT in terms of the stress tensor.

The case of the scalar field is analogous (a complete account will be presented elsewhere \cite{nuevo}). The massless limit has a local term in the modular Hamiltonian given again by (\ref{13}). Now the problem in the plane that gives the solution of the relevant eigenvalue problem is a function satisfying the Euclidean Klein-Gordon equation
\be
(-\nabla^2+m^2)S(\vec{x})=0\label{kg}
\ee
everywhere except at the cut $A$, where we impose the boundary condition
\be
S^{+}(x_1)=\lim_{x_2\rightarrow 0^+} S(x_1,x_2)= e^{2 \pi s} \lim_{x_2\rightarrow 0^-} S(x_1,x_2)=e^{2 \pi s}\, S^-(x_1)\,, \hspace{1cm} x_1\in A. \label{disc1}
\ee
 The relevant functions are the limit on the cut of the function and its time derivative, $S^+(x)$, $\partial_2 S^+(x)$. Because of the same reasons as above, we expect that the limit of large $|s|$ gives functions independent of the mass, for any fixed mass. Then the modular Hamiltonian has always the same leading local term.

In addition, the Appendices contain two explicit analytic calculations for the massive fermion field. In Appendix \ref{ap1} we have computed the modular Hamiltonian for one interval perturbatively in the mass to first order. We get some interesting information about non local terms, in particular there are "quasilocal" terms with fields in reflected positions, for example of the form $\sim m \int dx\,\frac{(x-a)(b-x)}{b-a} \psi^\dagger(x)\gamma^0\psi(b+a-x)$. We get a local term equivalent to the massless one but where $T_{00}(x)$ is replaced by the massive energy density. That is, the local term contains a (subleading for large energies) term $2\pi \int dx\, f(x)\,  m \psi^\dagger \gamma^0 \psi$, with the same coefficient as the one of $T_{00}$ in the massless case. In Appendix \ref{ap2} we show this last feature has to remain true non perturbatively in the mass. The reason is that causal propagation between different surfaces erases the local term of the mass and creates one from the leading local kinetic term. Then the coefficient of the mass term is always the same as the one of the kinetic term.

\begin{figure}[t]
  \begin{subfigure}[b]{0.5\linewidth}
    \centering
    \includegraphics[width=0.75\linewidth]{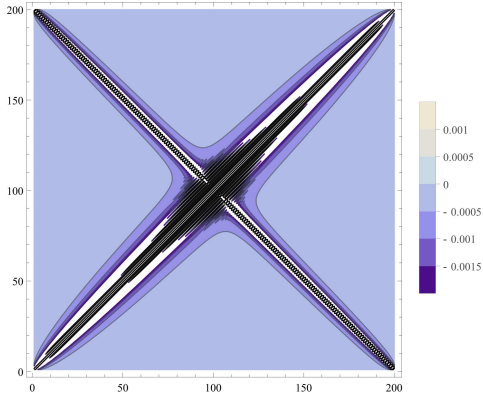}
    \caption{Matrix $M$ for $L=200$ and $mL=1$.}
    \label{M1:a}
    \vspace{4ex}
  \end{subfigure}
  \begin{subfigure}[b]{0.5\linewidth}
    \centering
    \includegraphics[width=0.75\linewidth]{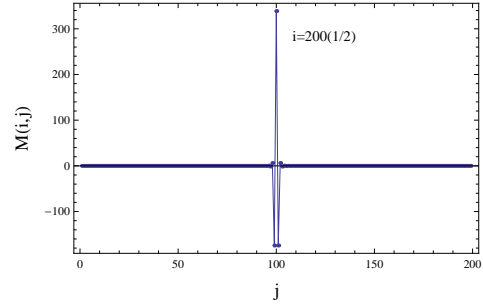}
    \caption{Second derivative of delta function in $M(i_0,j)$ for a segment of length $L=200$ and $i_0=\frac{1}{2}L$.}
    \label{M1:b}
    \vspace{4ex}
  \end{subfigure}
  \begin{subfigure}[b]{0.5\linewidth}
    \centering
    \includegraphics[width=0.75\linewidth]{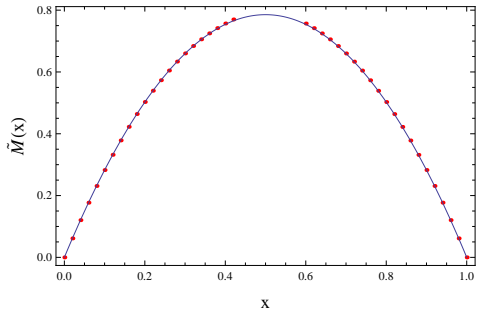}
    \caption{ The points correspond to the numerical result for the continuum limit of the coefficient $\tilde{M}$ of the $-\delta^{\prime\prime}$ kernel
     and the theoretical (solid line) result is  $\pi x(x-L)/L$ with $L=1$.}
    \label{M1:d}
  \end{subfigure}
  \caption{Scalar in a single interval. Kernel $M$ for a scalar field of mass $mL=1$.
  }
  \label{M1}
\end{figure}

\section{Free fields in $d=2$: Results from lattice calculations}

In this Section we test the results obtained in the previous section using techniques in the lattice for different configurations.
We calculate the modular Hamiltonian for massive scalar and fermion
fields on one and two intervals finding a perfect matching between the local parts of these modular Hamiltonians and the one in  (\ref{13}).
Our strategy is to calculate modular Hamiltonian kernels from the discrete two point correlators and from there, extract the local contribution $K_{loc}$. This would be straightforward but two subtleties make these calculations tricky.

The first one is that the modular Hamiltonian is roughly speaking proportional to inverse temperatures times energy. For high energy modes, and in particular for the local terms we are interested in, the contribution of the modular Hamiltonian is large, corresponding to the fact that these modes are little excited in the vacuum fluctuations. These large numbers in   $K$ come from logarithms of correlations functions that are vanishing small, and there are many modes that pile up to exponentially small eigenvalues of the correlations functions. These modes do not contribute much to the entanglement entropy and in consequence they do not need to be treated in detail in standard computations of entanglement entropy. However, they give important contributions to the modular Hamiltonian, specially for the local terms. As a result we are forced to use very high numerical precision to account for these modes, and the number of digits of precision needed  increases with the size of the region in the lattice and the mass. For lattices of up to $200$ point that we used we had to use around  $600$ digits to get consistent results. This large precision can be understood more quantitatively as follows. We have seen in the previous section that we have eigenvalues for the correlator (for the fermion for example) which pile up around $1$ as $1-e^{-2 \pi s}$ for large eigenvalue parameter $s$, see (\ref{eig}). Taking into account that $s$ acts as a momentum in a phase factor we can understand that for a lattice of $L$ points we are getting values of $s$ as large as $L/\epsilon$, or the number of lattice points. To get the right local term we need to get right the eigenvalues and eigenvectors for this large $s\sim L/\epsilon$. The corresponding eigenvalues differ from one by an exponentially small number. As an example for $L=200$ lattice points we get a rough indication of the correct number of digits needed $e^{-2 \pi\, 200}\sim 10^{-546}$. This also shows the difficulties in going to larger lattices.

The second subtlety comes from extracting the relevant delta function or its derivatives from the lattice results. We explain how this is achieved in more detail below.

\subsection{Scalar field}

\begin{figure}[t]
\centering
\leavevmode
\epsfysize=4cm
\epsfbox{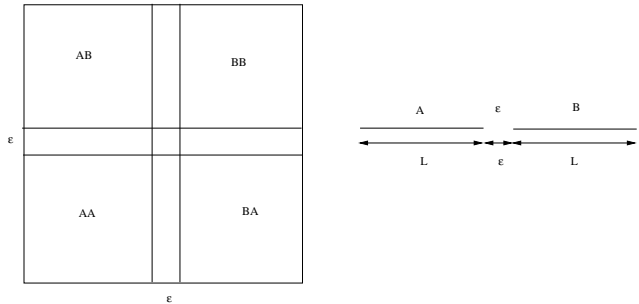}
\bigskip
\caption{The generic form of the kernel matrices for two intervals $A$, $B$, separated by a distance $\varepsilon$.}
\label{set2}
\end{figure}

Suppose we have a set of bosonic coordinates $\phi_i$ and its conjugate momenta $\pi_j$ with usual canonical commutation
relations $[\phi_i,\pi_j]=\delta_{ij}$ and $[\phi_i,\phi_j]=[\pi_i,\pi_j]=0$.
We define the vacuum correlators inside the region of interest $V$
\be
<\phi_i\phi_j>=X_{ij},~~~~~~~~~<\pi_i\pi_j>=P_{ij},~~~~~~~~~~~~<\phi_i\pi_j>=<\pi_j\phi_i>^*=\frac{i}{2}\delta_{ij}\label{correl}.
\ee
In one spatial dimension for a real massive scalar with Hamiltonian
\be
H=\frac12\sum_{n=\infty}^\infty\,(\pi_n^2+(\phi_{n+1}-\phi_n)^2+m^2\phi_n^2)
\ee
the two point correlators $X$ and $P$ are explicitly given by
\bea
<\phi_n\phi_m>&=&\int_{-\pi}^\pi dx\frac{e^{ix(m-n)}}{4\pi\sqrt{m^2+2(1-\cos x)}},\nn\\
<\pi_n\pi_m>&=&\int_{-\pi}^\pi dx \frac{1}{4\pi}e^{ix(m-n)}\sqrt{m^2+2(1-\cos x)}\label{latticecorrel}\,.
\eea

Moreover, in the free field case, it can be shown that the modular Hamiltonian $ K$ defined through the density matrix as
\be
\rho_V=c\,e^{-K}\,,
\ee
with $c$ such that $Tr(\rho)=1$, can be written in terms of the relevant two point correlators of the theory. For a review see \cite{HuertaReview}.
 We have
\be
 K=\sum\left(M_{ij}\phi_i\phi_j+N_{ij}\pi_i\pi_j\right)\,,
\label{latticeham}
\ee
with
\be
M=P.L \,,\hspace{.7cm}
N=L.X\,,
\label{MN}\hspace{.7cm} L=\frac{1}{2C}\log\left(\frac{C+\frac12}{C-\frac12}\right)\,,
\ee
 where $C=\sqrt{XP}$.
\subsubsection{One interval}
Since we are interested in the continuum limit of the local modular Hamiltonian, we calculate (\ref{latticecorrel}) and (\ref{latticeham})
for different masses $m_{\lambda}$ and interval lengths $L_{\lambda}$, keeping $m_{\lambda}L_{\lambda}$ fixed, to finally read the local modular Hamiltonian from the
limit  $L_{\lambda}\rightarrow\infty$ of the relevant kernels.
More specifically, for a given $L\le L_{max}$, where $L_{max}=200$ is fixed by the total lattice size, we first calculate the correlators $X_{ij}$, $P_{ij}$, with $i, j \leq L$, and mass $m = \textrm{const}/L$ to
finally evaluate the modular hamiltonian  from (\ref{MN}).
We repeat the calculation for different lengths $L_{\lambda} = 50\lambda$ with $\lambda = 1, 2, 3, 4$ keeping $m_{\lambda}L_{\lambda}=(mL)_{\textrm{continuum}}$ fixed.

\begin{figure}[t]
  \begin{subfigure}[b]{0.49\linewidth}
    \centering
    \includegraphics[width=0.7\linewidth]{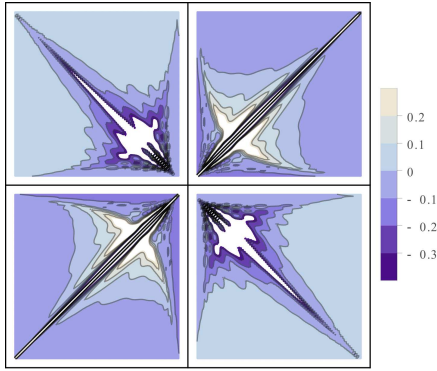}
    \caption{Matrix $N$, for $N_{\textrm{max}}=200$, each interval of \newline length $L=90$ and separation $\varepsilon=20$.
    \vspace{1.cm}}
    \label{N2:a}
    \vspace{4ex}
  \end{subfigure}
  \begin{subfigure}[b]{0.5\linewidth}
    \centering
    \includegraphics[width=0.75\linewidth]{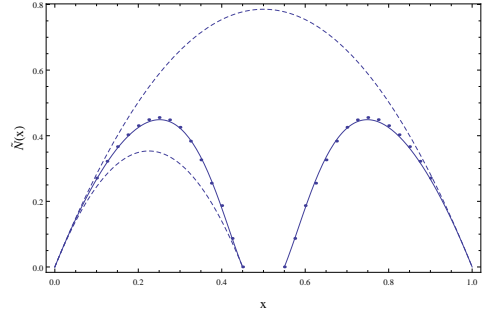}
    \caption{Comparison of the numerical continuum limit for the coefficient of the local term proportional to a delta function (points) with the expected value, formula (\ref{quant}) (solid line). The lower dashed curve corresponds to the corresponding coefficient for a single interval $L$ and the upper dashed curve to the coefficient for a single interval of size $N_{\textrm{max}}$.}
    \label{N2:b}
    \vspace{4ex}
  \end{subfigure}
   \caption{Kernel $N$ for a scalar in two intervals. The total size of the lattice is $N_{\textrm{max}}=2L+\varepsilon$, the mass $m N_{\textrm{max}}=1$, and $\varepsilon=N_{\textrm{max}}/10$.}
  \label{N2}
\end{figure}

In figure (\ref{N1:a}) it is shown what we obtain for the kernel $N^{\lambda=4}$. In this example, we can see that $N^{\lambda}$ is {\sl almost} diagonal.
In fact, if we plot $N^{\lambda}(a,j)$ for a fixed $a$ as a function of
 $j$, we find that the main contribution comes from the diagonal but there are also smaller ones from first and second
neighbours $N(a, a\pm 1,2)$. This is shown in figure (\ref{N1:b})
for the case $a=\frac{3}{5}L$ and the maximal size of the lattice $\lambda=4$ ($200$ points). Here we list some numbers for the $\lambda=4$ case
\bea
N((3/5) L,(3/5)L)&=&159.11838\,,\\
N((3/5)L,(3/5)L-1)&=&-2.22265\,,\\
N((3/5)L,(3/5)L+1)&=&-2.16739\,.
\eea
On the other hand, according to our anzats for one interval
\be
K_{\textrm{loc}}= 2\pi\int_0^L dx\, \frac{x(L-x)}{L} T_{00}(x) \,,
\ee
it should be in the continuum
\begin{equation}
N_{\textrm{loc}}(x,y)= \pi f(x) \delta(x-y)\,,
\label{nn}
\end{equation}
with $f(x)=\frac{x(L-x)}{L}$.

Naively, the delta function in (\ref{nn}) should be associated to the diagonal $N^{\lambda}_{ii}$.
Instead, we find a better approximation in the lattice to the continuum delta with
the correct scaling $\tilde{N}=N(i, i-1) + N(i, i) + N(i, i + 1)$. This quantity in the lattice grows linearly with $\lambda$ as it must be from the delta function contribution to (\ref{nn}). There should be also some other next neighbours contributing to the delta function but we find their contribution is small. These contributions also get mixed with non local terms, but the non local terms do not scale with $\lambda$ and we can select the local term doing a fit in $\lambda$.

We can formalize this approximation to the delta function as follows. This will be useful later when dealing with delta derivative kernels.  Let us evaluate $ \sum_{j=i-1}^{i+1}b_j g(x_j)$ with $g(x)$
a test function and $x_i=i \epsilon$ the continuum coordinate of the point $i$, with $\epsilon$ the lattice spacing.
Then,
\begin{eqnarray}
\sum_{j=i-1}^{i+1}b_j g(x_j) = b_{i-1} g(x_{i-1})+b_i g(x_i)+b_{i+1} g(x_{i+1})\nonumber\\
 =(b_{i-1}+b_i+b_{i+1}) g(x_{i})+(b_{i+1}-b_{i-1}) g^\prime(x_i) \epsilon+ {\cal O}(\epsilon^2)\,.\label{smsm}
\eea
Hence, the sum of the lattice values corresponds to the coefficient of a delta function. If these values go to zero fast enough outside the diagonal is enough to keep first neighbours as we do here.

\begin{figure}[t]
  \begin{subfigure}[b]{0.57\linewidth}
    \centering
    \includegraphics[width=0.65\linewidth]{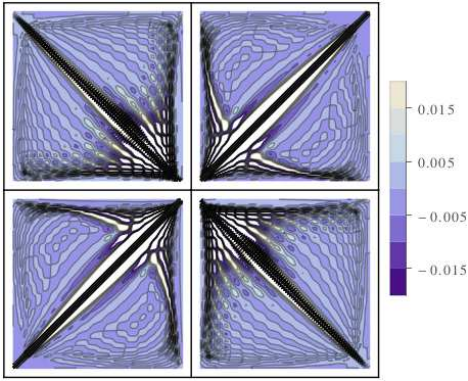}
    \caption{Matrix $M$, for $N_{max}=200$, each interval of length \newline  $L=90$ and separation $\varepsilon=20$.\vspace{.2cm}}
    \label{M2:a}
    \vspace{4ex}
  \end{subfigure}
  \begin{subfigure}[b]{0.4\linewidth}
    \centering
    \includegraphics[width=1.05\linewidth]{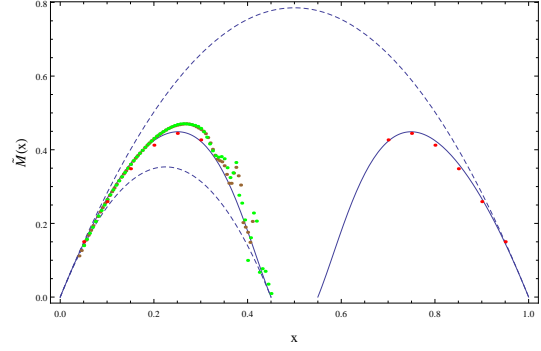}
    \caption{Comparison of the continuum limit (red points) to
     the maximal size for $N_{\textrm{max}}=200$ (green points). The solid and dashed curves are as in figure \ref{N2:b}.  }
    \label{M2:b}
    \vspace{4ex}
  \end{subfigure}
   \caption{Kernel $M$ for a scalar in two intervals with equal size $L$ separated by a distance $\varepsilon$. The mass $m N_{\textrm{max}}=1$ and $\varepsilon=N_{\textrm{max}}/10$, with $N_\textrm{Max}=2L+\varepsilon$.}
  \label{M2}
\end{figure}

We show in fig(\ref{N1:c}) the results for the case $mL =3$ and $L=25\lambda$ and $1\leq \lambda\leq 8$.
Once the identification of $(N^{\lambda}_{ii}+N^{\lambda}_{ii+1}+N^{\lambda}_{ii-1})$ with the coefficient of the delta function is done, we look for the continuum limit. For that, we fit the pairs $(\lambda, N^{\lambda}_{ii}+N^{\lambda}_{ii+1}+N^{\lambda}_{ii-1})$
for each point $ \lambda x_i$, $i<25$ with the function $a_0 + a_1 (25 \lambda) + a_{-1} (25 \lambda)^{-1}$. The continuum limit of the coefficient of the delta function (for $L=1$) corresponds to the
linear coefficient $a_1$, that we call
$ \tilde{N}_{x}=a_1(x)$.

The result $a_1$ (scaled to $L=1$) is shown in figure (\ref{N1:d}) for the case $m L=1$. This agrees very well with the prediction. For example, for the middle point $x=\frac{1}{2}$ we obtain $0.7826$
compared to the expected value $\frac{\pi}{4}=0.7853$.

The same analysis has been done for the kernel $M$ (fig. \ref{M1}). In fig \ref{M1:a} we show what we obtain for the case $L=200$ and $mL=1$.
According to our anzats, $M^{loc}$ should be
\begin{equation}
M^{loc}(x,y)=-\pi f(x) \delta^{\prime\prime}(x-y)\,
\label{mm}
\end{equation}
with $f(x)= \frac{x(L-x)}{L}$.
In this case, the identification of the discrete version of $\delta^{\prime\prime}(x-y)$ follows from an argument analogous to the one used in the identification of the $\delta(x-y)$.

For a test function $g(x)$ let us evaluate again (\ref{smsm}) but up to order $\epsilon^2$, and keeping up to five neighbours
\be
 \sum_{j=i-1}^{i+1}b_j g(x_j) = \left(\sum_{j=i-5}^{i+5} b_j \right) g(x_i) +\left(\sum_{j=i-5}^{i+5} (i-j) b_j\right) g^\prime(x_i) \epsilon +\left(\sum_{j=i-5}^{i+5} b_j \frac{(i-j)^2}{2}\right) g^{\prime\prime}(x_i) \epsilon^2 +\hdots \label{pol}
\ee
Then, to select the coefficient of the second derivative of the delta function we have to sum the lattice values around the diagonal with certain weights dictated by the last term in (\ref{pol}), $
\tilde{M}=-(M(x,x+1)+4M(x,x+2)+9M(x,x+3)+16M(x,x+4)+25M(x,x+5))$ where we use that the data is highly symmetric around the diagonal.  We use a sum up to $5$ neighbours because we find this improves the result. While the data gets small fast outside the diagonal, still the coefficients in the sum increase, and we have to take these terms into account.

\begin{figure}[t]
  \begin{subfigure}[b]{0.5\linewidth}
    \centering
    \includegraphics[width=0.7\linewidth]{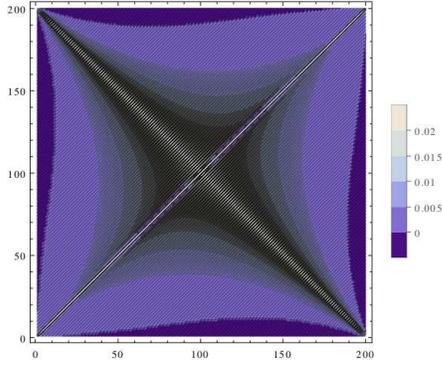}
    \caption{Matrix $H_{0}$ for $L=200$. \vspace{.6cm}}
    \label{h01:a}
    \vspace{4ex}
  \end{subfigure}
  \begin{subfigure}[b]{0.5\linewidth}
    \centering
    \includegraphics[width=0.85\linewidth]{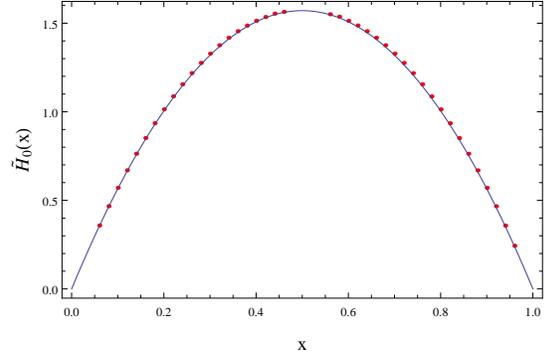}
    \caption{Comparison of the lattice continuous limit with the theoretic result. We have used up to second neighbours to extract the delta function.}
    \label{h01:b}
    \vspace{4ex}
  \end{subfigure}
    \begin{subfigure}[b]{0.5\linewidth}
    \centering
    \includegraphics[width=0.85\linewidth]{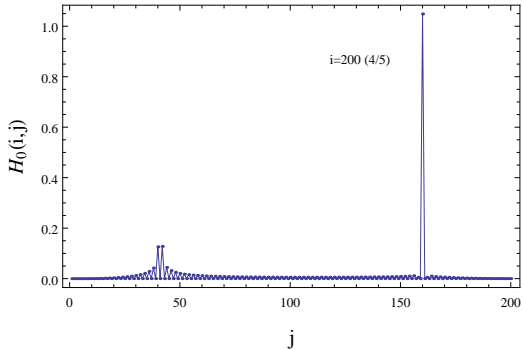}
    \caption{
 Delta function
in $H_{0}(i_0,j)$ for a segment of length $L=200$. There is a
term in the anti-diagonal which is also seen perturbatively as shown in Appendix \ref{ap1}.}
    \label{h01:c}
  \end{subfigure}
  \caption{Fermion in a single interval. Kernel $H_{0}$ for a massive fermion field of mass $mL=1$.
}
  \label{h01}
\end{figure}

For getting the continuum limit we use $L=50 \lambda$, $\lambda=1,2,3,4$. The numerical continuum limit corresponds to $a_1$ in the fit $a_0+a_1 (50 \lambda)+a_{-1}(50\lambda)^{-1}$ for $\tilde{M}$. This is plotted in figure (\ref{M1:d}).

\begin{figure}[t]
  \begin{subfigure}[b]{0.45\linewidth}
    \centering
    \includegraphics[width=0.75\linewidth]{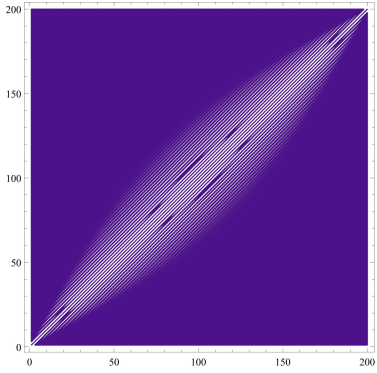}
    \caption{Matrix $H_{1}$ with $L=200$. \vspace{.8cm}}
    \label{h11:a}
    \vspace{4ex}
  \end{subfigure}
  \begin{subfigure}[b]{0.6\linewidth}
    \centering
    \includegraphics[width=0.7\linewidth]{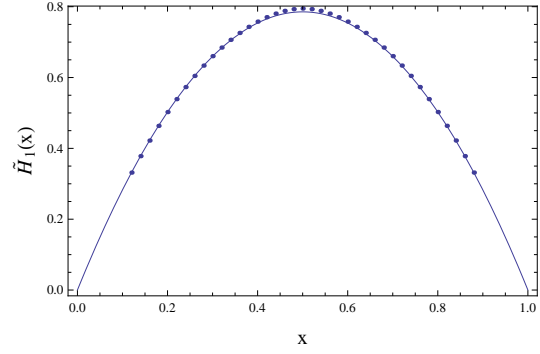}
    \caption{Comparison of the numerical continuum limit of the coefficient of the derivative of the delta function and the theoretical result.  We used up to $5$ neighbours to extract the local term.}
    \label{h11:b}
    \vspace{4ex}
  \end{subfigure}
    \begin{subfigure}[b]{0.6\linewidth}
    \centering
    \includegraphics[width=0.75\linewidth]{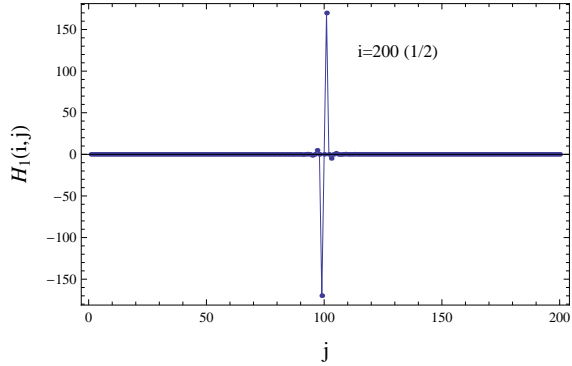}
    \caption{Numerical shape of the delta prime function, for $L=200$.}
    \label{h11:c}
  \end{subfigure}
  \caption{Fermion in a single interval. Kernel $H_{1}$ for a massive fermion field of mass $mL=1$.
}
  \label{h11}
\end{figure}

Let us note that in fig.(\ref{M1:d}) the continuum limit for middle points is not included. The reason is that in this region,
as seen in fig.(\ref{M1:a}), there is also an anti-diagonal non local contribution. This mixes with the local one and introduces unwanted corrections to $\tilde{M}$
that do not correspond to the $\delta^{\prime\prime}$ we are looking for.

\subsubsection{Two intervals}

We also study the case of two intervals shown in fig. \ref{set2}. In this case, we take sets
with $N_{max}=2L+\varepsilon=40\lambda$ with $\lambda=1,2,3,4,5$, $\varepsilon=N_{max}/10$ and $m=1/N_{max}$.
We repeat the previous analysis for the kernels $N$ shown in figure \ref{N2} and $M$ in figure \ref{M2}.
In these cases, we have used a very high numerical precision. The correlators are calculated with $800$ digits. According to our experience,
the precision we have used for one interval that was $360$ digits for $mL=1$ and $600$ digits for $mL=3$, was not enough in the present case. As before, we find the local kernels match with the expectations (\ref{13}),
\bea
N^{loc}(x,y)&=& \pi f(x) \delta(x-y)\,,\\
M^{loc}(x,y)&=&- \pi f(x) \delta^{\prime\prime}(x-y)\,,
\label{mm1}
\eea
with
 \be
 f(x)=\frac{x (\frac{N_{max} -\varepsilon}{2}-x)(\frac{N_{max} + \varepsilon}{2}-x)(N_{max} - x)}{( N_{max}-\varepsilon)
(\frac{\varepsilon N_{max}}{4} +
(x -\frac{N_{max}}{2})^2)}\,.\label{quant}
\ee
In the kernel $M$, we find the reading of the $\delta^{\prime\prime}$ is much more difficult than in the one interval case. Again,
this is due to the non local contributions as seen in fig. \ref{M2:b} where we have included the $\tilde{M}_{\lambda}$ for $\lambda=3,4$.
In the plot, one sees that for the points in the first (second) interval closer to the second (first) one, the non local contributions are stronger
and strongly deform the local $\delta^{\prime\prime}$ for the sizes up to $200$ points we are considering.

\subsection{Massive fermion fields}

For fermion fields, the modular Hamiltonian kernel $H$ can be written in terms of the non-vanishing two point correlator \cite{HuertaReview}
\begin{equation}
H=-\log \left(C^{-1}-1\right)\,,
\label{modfer}
\end{equation}
where
\begin{equation}
C_{ij}=\langle \psi_i \psi_j^{\dagger}\rangle\,.\label{correfer}
\end{equation}
The correlator (\ref{correfer}) in the lattice is given by
\begin{equation}
C_{jk}=\frac{1}{2}\delta_{jk}-\int_{-\pi}^{\pi} dx \frac{m\gamma^0+\sin(x)\gamma^0 \gamma^1}{4\pi\sqrt{m^2+\sin^2(x)}}e^{-i x (j-k)}\,.
\end{equation}

\begin{figure}[t]
  \begin{subfigure}[b]{0.5\linewidth}
    \centering
    \includegraphics[width=0.75\linewidth]{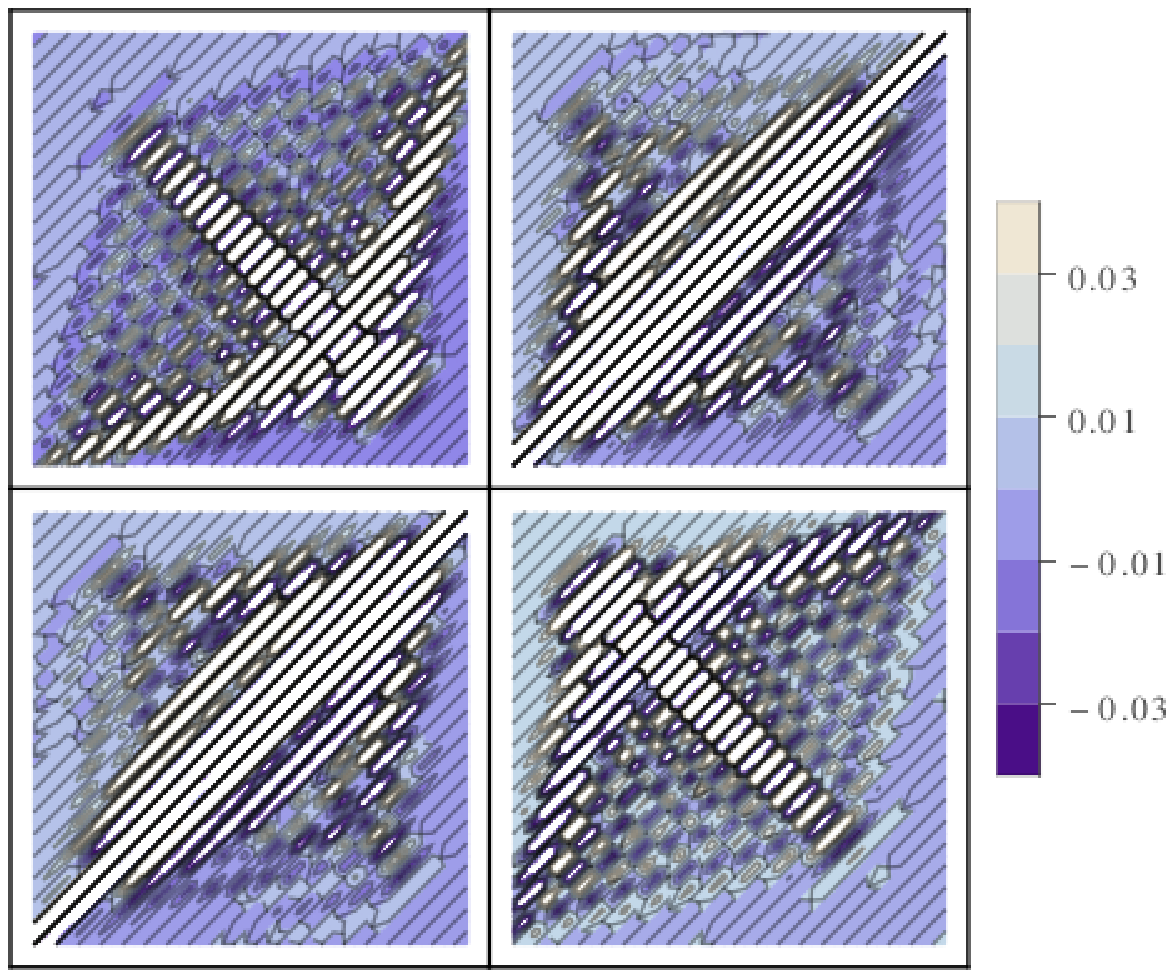}
    \caption{Matrix $H_1$, for $N_{\textrm{max}}=200$, each interval \newline of length $L=80$ and separation $\varepsilon=40$.}
    \label{h12:a}
    \vspace{4ex}
  \end{subfigure}
  \begin{subfigure}[b]{0.5\linewidth}
    \centering
    \includegraphics[width=0.8\linewidth]{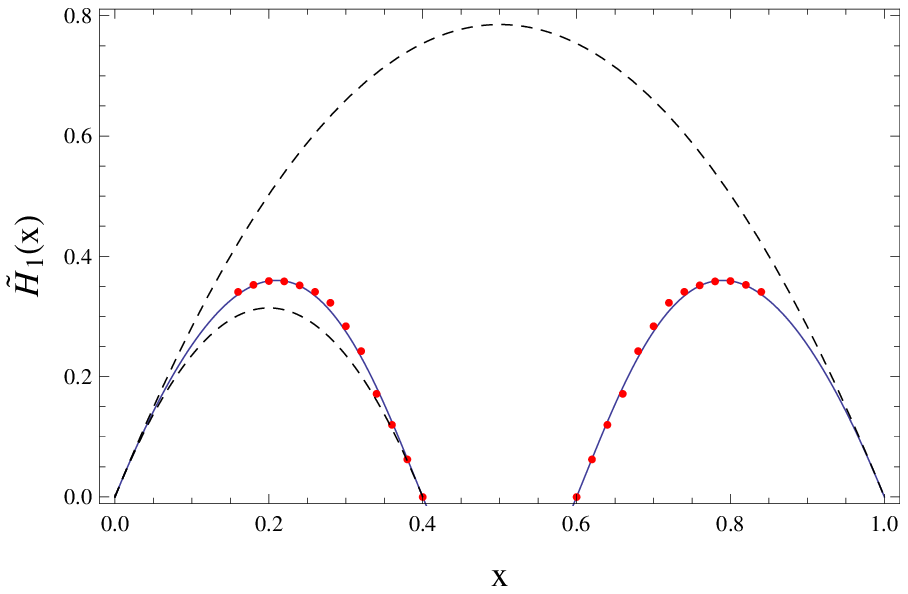}
    \caption{Comparison of the numerical continuum limit of the coefficient of the delta prime and the theoretical result.
We have used up to $7$ neighbours to extract the local term.}
    \label{h12:b}
    \vspace{4ex}
  \end{subfigure}
   \caption{Fermion in two intervals. Kernel $H_1$ for a massive fermion field of mass $m N_{\textrm{max}}=1$ and $\varepsilon=N_{\textrm{max}}/5$.}
\label{h12}
\end{figure}

\begin{figure}[t]
  \begin{subfigure}[b]{0.5\linewidth}
    \centering
    \includegraphics[width=0.72\linewidth]{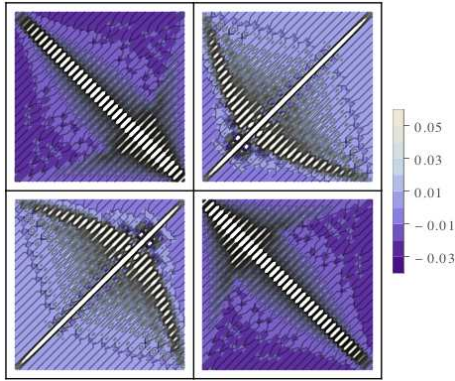}
    \caption{Matrix $H_0$, for $N_{\textrm{max}}=200$, each interval\newline of length $L=80$ and separation $\varepsilon=40$.}
    \label{h02:a}
    \vspace{4ex}
  \end{subfigure}
  \begin{subfigure}[b]{0.53\linewidth}
    \centering
    \includegraphics[width=0.7\linewidth]{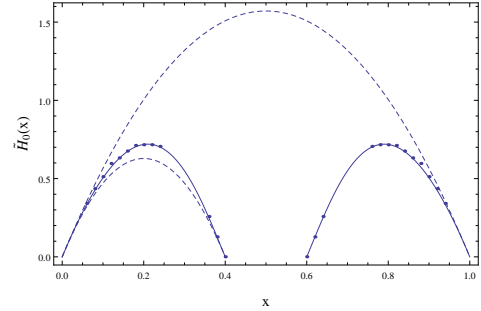}
    \caption{ The continuum limit of the numerical data compared with the theoretical result. We have used a fit with powers of $\lambda$ up to $\lambda^{-2}$ to get the continuum limit and second neighbours to extract the local term.}
    \label{fh02:b}
    \vspace{4ex}
  \end{subfigure}
  \begin{subfigure}[b]{0.53\linewidth}
    \centering
    \includegraphics[width=0.7\linewidth]{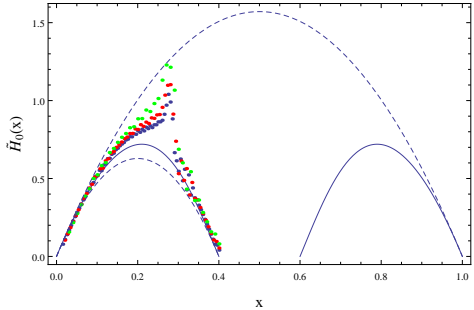}
    \caption{The points correspond to different $N_{\textrm{max}}=50\lambda$ with $\lambda=2,3,4$.
The solid line corresponds to the theoretical function (\ref{quant}).}
    \label{h02:b}
    \vspace{4ex}
  \end{subfigure}
   \caption{Fermion in two intervals. Kernel $H_0$ for a massive fermion field of mass $m N_{max}=1$ and $\varepsilon=N_{\textrm{max}}/5$.}
  \label{h02}
\end{figure}

\subsubsection{One interval}
In this case, for a set of length $L$, the correlator and the modular Hamiltonian are matrices of $2L\times2L$, since for each pair $(i,j)$,
$C_{ij}$ and $H_{ij}$ are  $2\times2$ matrices.
As before, we calculate (\ref{correfer}) and (\ref{modfer}) for different masses $m_{\lambda}$ and segment lengths $L_{\lambda}=50 \lambda$ with $\lambda =1,2,3,4$
such that $m=1/L$.
In figure (\ref{h11:a}) and (\ref{h01:a}) is shown what we obtain for ${H_{0}}(i,j)$ and ${H_{1}}(i,j)$ in (\ref{modfer})  for $L=200$, where we have defined
\begin{equation}
H=H_{1} \gamma^0 \gamma^1+H_{0}\gamma^0\,.
\end{equation}
Here $H_{0}$ and $H_{1}$ are $L\times L$ matrices.
According to our results it should be these kernels contain local terms in the continuum limit
\begin{equation}
H_{0}(x,y)\sim \pi f(x) \,2\, m\, \delta(x-y) \,,
\label{h0}
\end{equation}
and
\begin{equation}
H_{1}(x,y)\sim \pi f(x) \,2\, \, \delta^{\prime}(x-y) \,,
\label{h01b}
\end{equation}
with $f(x)=\frac{x(L-x)}{L}$.

In the kernel $H_{0}$, the identification of the delta function in the lattice is as above:
we consider the quantity $\tilde{H}=H_{0}(i,i)+H_{0}(i,i+1)+H_{0}(i,i-1)$ with first neighbours. In this case
the scaling with $L$ is different from the one of the scalar kernel $N$. Here the presence of the mass in (\ref{h0})
makes $\tilde{H}$ independent of $L$ (we are taking $m=1/L$).

On the other hand, in the kernel $H_{1}$, we again have to find a quantity which scales linearly with $L$.
From the same argument already used for the delta and second derivative delta functions, we find
$\tilde{H_1}=(H_{1}(i,i+1)-H_{1}(i,i-1))+3(H_{1}(i,i+3)-H_{0}(i,i-3))+...$.

For $H_0$, shown in fig. \ref{h01}, as in the scalar kernel $M$, there is a non local (antidiagonal) contribution. Because of this, we skip
the continuum limit of the middle points. In any case the data shows very good agreement with the expectations.

\subsubsection{Two intervals}
For the two intervals case, we take $N_{max}=2L+\epsilon=50\lambda$, $\epsilon=N_{max}/5$, $m=1/N_{max}$, $\lambda=1,2,3,4$.
The kernels $H_{0}$ and $H_{1}$ are shown in figures (\ref{h02}) and (\ref{h12}).
Here again we find the non local contributions in $H_0$ makes difficult the calculation of the continuum limit for the points
in the first (second) interval close to the second (first) interval. However, again we find overall agreement with expectations.

\section{Discussion}

The modular Hamiltonian of the Rindler wedge is local and proportional to the energy density. This locality cannot be generalized to other regions in general QFT; the corresponding modular Hamiltonians must necessarily contain non local terms.  We have proposed that the right path to generalize the Rindler result is to look at contributions to the relative entropy with localized excitations. This allows us to define local temperatures analogous to Unruh temperature in Rindler space. We have seen that local temperatures in null directions have some degree of universality. In particular, they are bounded by universal geometric quantities.  We have shown these temperatures are the same for $d=2$ free scalar and fermions independently of the mass. The mass should not alter the null temperatures for free fields in higher dimensions either.

The strongest conjecture one could formulate given our present understanding is the following. In any dimensions null temperatures would be uniquely defined by the geometry and be universal across all QFT. Further, we could also think there is no range of possible temperatures in null directions. We have shown this last statement holds for free scalars.

This proposal is a universal connection between temperature and geometry through vacuum entanglement in high energy modes. It tells of an imprint of the geometry on the high energy tail of the reduced density matrices, exactly as it happens in the Hawking radiation from a black hole. If the conjecture is correct, the peculiarity of the Rindler wedge for all QFT, and of spheres in CFT, would be only that the non local terms exactly cancel in these cases, while the null temperatures would extend the geometric universality to any region.

It is interesting to note that thinking in the space of all QFT, the statement that there is a unique null temperature for each theory, that is, the maximum and minimum null temperatures agree, implies the proposed universality. If the null temperatures are not universal, combining independent theories we would obtain another theory with a range of possible null temperatures.

These null temperatures could come from the projections of an inverse temperature vector in $d=2$, and hence correspond to a local term involving the stress tensor in this case, but this is not possible for $d>2$ and generic Cauchy surfaces. For $d>2$ there must be contributions other than the stress tensor to the modular Hamiltonian, but they should still scale as the energy for localized excitations.

There are several questions we left to future research. Perhaps the natural next step is to understand the case of interacting CFT in $d=2$. For $d>2$ we know the local temperatures exist, but in interacting theories we do not know the structure of the operators in $K$ that gives place to these contributions.

For $d=2$ we argued that a simple analytic continuation in the complex plane of the propagation equations of the null temperatures gives place to the right null temperatures found for free models. It would be interesting to generalize this purely geometric argument to more dimensions. We need an analogous equation that extends holomorphically the statement that null temperatures are functions of null rays, and then impose the vanishing of the modular Hamiltonian on the boundary of the region at the same rate as in Rindler space, as a boundary condition. Perhaps this question can be explored using twistor techniques.

Another interesting question is if some of these structures survive for states different from the vacuum. In principle the local temperatures defined for the limit of large energies should not depend on low energy excitations above the vacuum, such as a single particle state or a coherent state in a free theory.
Hence, there should be a large number of states sharing the same null temperatures.  On the other hand, some mild perturbations of the state, but containing an exponential tail of large energy excitations would modify the temperatures in an important way, specially far from the boundaries. This is the case of a thermal state. For the free massless models in $d=2$, and for a general $d=2$ CFT in an interval \cite{exa1}, the thermal case can be obtained by conformal transformations. It is clear that the local temperatures are given by solving a problem about a holomorphic function with poles at the boundary of the region analogous to the one solved in Section \ref{argument}, but where now the space where this function lives is not the complex plane but the cylinder. We wonder for what class of states the local temperatures could be similarly codified in a geometry. There are other states like $\rho\sim e^{-\gamma \sqrt{H}}$ that are "superthermal" and local temperatures cannot be defined for them.

\section*{Acknowledgements}
We thank John Cardy and Erik Tonni for discussions, and Robert Myers for useful comments on the manuscript. This work was partially supported by CONICET, CNEA and Universidad Nacional de Cuyo, Argentina. H.C. acknowledges support from an It From Qubit grant of the Simons Foundation.

\appendix

\section{Expansion of the modular hamiltonian for small mass for a $d=2$ Dirac field}
\label{ap1}

In this Appendix we compute the modular Hamiltonian of a Dirac field in $d=2$ for one interval of length $L$ to first order in the mass. The  modular Hamiltonian is quadratic in the fields and has the form
\be
K=\int_0^L dx\, dy\, \psi^\dagger(x) H(x,y) \psi(y)\,.
\ee
with the kernel $H(x,y)$ given by \cite{HuertaReview}
\be
H=-\int_{1/2}^\infty d\beta\, (R(\beta)+R(-\beta))\,,
\ee
in terms of the resolvent
\be
R(\beta)=(C-1/2+\beta)^{-1}\,,
\ee
where $C(x,y)=\langle 0|\psi(x)\psi^\dagger(y)|0\rangle$ is the correlator kernel in the interval of size $L$. Expanding $C$ to first order in the mass we have
\be
R(\beta)=R^0(\beta)-R^0(\beta)\,\delta C \,R^0(\beta)+...\label{23}
\ee
with \cite{fermion}
\bea
R^0(\beta)(x,y)&=&\int_{-\infty}^\infty ds\, \psi_s(x) M(\beta,s) \psi^*_s(y),~~M(\beta,s)=\left(\beta {\bf 1}-\tanh(\pi s)\frac{\gamma^3}{2}\right)^{-1}\,,\\
\psi_s(x)&=& \frac{L^{1/2}}{(2\pi)^{1/2}\sqrt{x (L-x)}}e^{-i s z(x)},~~z(x)=\log\left(\frac{x}{(L-x)}\right)\,,\\
\delta C(x,y)&=&-\frac{m}{2\pi}\left(\gamma_E-\log(2)+\log(m|x-y|)\right)\gamma^0\,.
\eea
Here $\gamma^3=\gamma^0\gamma^1$, with $\gamma^\mu$ the Dirac matrices, and $\gamma_E$ is the Euler constant. The eigenvectors are normalized as in (\ref{normi}).

\subsection*{The zero order}
At zeroth order in a mass expansion we have that
\be
\int_{1/2}^\infty d\beta\, (M(\beta,s)+M(-\beta,s)=2 \pi s \gamma^3\,.
\ee
and then
\bea
H_0&=&-\gamma^3\int_{-\infty}^\infty ds\,\frac{ s L\,e^{-i s (z(x)-z(y))}}{(x(L-x)y(L-y))^{1/2}} =\frac{2\pi i L \gamma^3}{(x(L-x)y(L-y))^{1/2}} \delta^{\prime}(z(y)-z(x))\nonumber \\
&=& i 2\pi \gamma^3 \left(\frac{x(L-x)}{L}\delta^{\prime}(x-y)+\frac{(L-2 x)}{2L} \delta(x-y)\right)\,.
\eea
The term proportional to the delta function is just such that the whole kernel is Hermitian. Taking into account that the Dirac Hamiltonian writes
\be
i\alpha_x \partial_x=i\gamma^3\partial_x \,,
\ee
this gives the zeroth order term for the modular Hamiltonian
\be
K_0=\int_0^L dx\, 2\pi \frac{x(L-x)}{L} T_{00}(x)\,,\label{espre}
\ee
with $T_{00}=(i/2) \psi^\dagger \gamma^3 \stackrel{\leftrightarrow}{\partial}_x \psi$.

\subsection*{The first order}
Now we want to compute the first order correction to the local part of the modular Hamiltonian due to the mass of the fermion. In order to do that we integrate in $\beta$ first
\be
\int_{1/2}^\infty d\beta\, (M(\beta,s)\gamma^0 M(\beta,s^\prime)+M(-\beta,s)\gamma^0 M(-\beta,s^\prime))=4\pi \gamma^0 (s+s^\prime) \frac{\cosh(\pi s)\cosh(\pi s^\prime)}{\sinh(\pi(s+s^\prime))}\,.
\ee
For doing the intermediate integral in (\ref{23}) we separate the contribution of $\delta C$ in two parts, constant, and proportional to $\log|x-y|$. Let us call the constant term
\be
k=-\frac{m}{2\pi}\left(\gamma_E-\log(2)+\log(m)\right)\,,
\ee
and the corresponding contribution to $H$ we call $H_{1,0}$.
Using \cite{fermion}
\be
\int_0^L dx\, \psi_s(x)=\left(\frac{\pi}{2}\right)^{1/2} L^{1/2} \textrm{sech}(\pi s)
\ee
we get
\be
H_{1,0}=k L^2 \pi \gamma^0 \int_{-\infty}^\infty ds\,\int_{-\infty}^\infty ds^\prime\,\frac{e^{-i (s z(x)- s^\prime z(y))}}{(x(L-x)y(L-y))^{1/2}}   \frac{(s+s^\prime)}{\sinh(\pi(s+s^\prime))}\,.
\ee
Changing variables to $u=s+s^\prime$ and $v=s-s^\prime$ we obtain
\bea
H_{1,0}=k L^2 \pi^2  \gamma^0 \frac{\delta(z(x)+z(y))}{(x(L-x)y(L-y))^{1/2}}  \int_{-\infty}^\infty du\, \frac{u\, e^{-i \frac{u}{2}(z(x)-z(y))}}{\sinh(\pi u)}
=4 k  \pi^2  \gamma^0 \frac{x y}{L}\,\,\delta(x+y-L)\,.\label{h10}
\eea
This is a curious non local term. It is not completely non local since it mixes only $x$ with $L-x$. It modifies the form of the Rindler modular Hamiltonian near the boundary, but non locally with the other boundary.

Now, we are going to study the contribution to the modular hamiltonian proportional to $m \log|x-y|$ and we will name this as $H_{1,1}$. We have to compute
\be
H_{1,1}=-2 m \gamma^0 \int ds \int ds'\,(s+s^\prime) \frac{\cosh(\pi s)\cosh(\pi s^\prime)}{\sinh(\pi(s+s^\prime))}\psi_s(x)\psi^*_{s'}(y)\int dx'\int dy'\psi_s^*(x')\log|x'-y'|\psi_{s'}(y').\label{h11a}
\ee
In order to do the integrals we define a function
\be
F(x')=\int dy'\log|x'-y'|\psi_{s'}(y').\label{wrt12}
\ee
and we use the fact that \cite{fermion}
\be
F'(x')=\int dy^\prime\,\frac{1}{x^\prime-y^\prime}\psi_{s^\prime}(y^\prime)=i\pi\tanh{\pi s^\prime}\psi_{s^\prime}(x^\prime),
\ee
and then
\be
F(x^\prime)=i \pi \tanh(\pi s^\prime)\int_0^{x^\prime} dx\,\psi_{s^\prime}(x)+A\,,
\ee
where $A$ is a constant of integration that we can fix using the value of the integral on $y'$ for $x'=0$ in \eqref{wrt12}
\be
A=\int dy'\log|-y'|\psi_{s'}(y')=\sqrt{\frac{L\pi }{2}}\,\text{sech}\left(\pi s'\right) \left(\log (L)+H_{-i s'-\frac{1}{2}}\right).
\ee
Here $H_{-i s'-\frac{1}{2}}$ the Harmonic number function. With this we arrive to
\be
F(x')=\sqrt{\frac{L\pi }{2}}\,\text{sech}\left(\pi s'\right) \left(\log (L)+H_{-i s'-\frac{1}{2}}\right)-\frac{\sqrt{2 \pi } \tanh \left(\pi  s'\right) \left(x'\right)^{\frac{1}{2}-i s'} \left(L-x'\right)^{\frac{1}{2}+i s'} \,
   _2F_1\left(1,1;\frac{3}{2}-i s';\frac{x'}{L}\right)}{\sqrt{L} \left(2 s'+i\right)}\,,
\ee
and the correction to the modular Hamiltonian kernel can be written as
\be
H_{1,1}=-2m\gamma^0 \int ds \int ds'\,(s+s^\prime)\frac{\cosh(\pi s)\cosh(\pi s^\prime)}{\sinh(\pi(s+s^\prime))}\psi_s(x)\psi^*_{s'}(y)\int dx'F(x')\psi_s^*(x').
\label{h11dos}
\ee
Now, we perform the integral in $x'$ and obtain the anti-symmetrized expression in the variables $s, s'$
\bea
\int dx'F(x')\psi^*_s(x')&=&\frac{1}{4} \pi\,  L\, \text{csch}\left(\pi  \left(s-s'\right)\right) \left[\tanh (\pi  s) \left(H_{-i s'-\frac{1}{2}}+H_{i s'-\frac{1}{2}}+2 \log
   (L)\right)\right.\nn\\&&\left.-\tanh \left(\pi  s'\right) \left(H_{-i s-\frac{1}{2}}+H_{i s-\frac{1}{2}}+2 \log (L)\right)\right].
   \eea
Note that the terms proportional to $\log(L)$ gives integrals like those in \eqref{h10} and then the result is going to be proportional to $\delta(x+y-L)$,
\be
-m L \pi \log(L)\int ds ds'\psi_s(x)\psi_s'^*(y)\frac{s+s'}{\sinh(\pi(s+s'))}=-2\pi \,m \log(L)\frac{x y}{L}\delta(x+y-L).\ee

The terms involving the harmonic functions can be integrated using the the change of variables $u=s+s', v=s-s'$. With this, the expression to study is
\bea
&&\frac{m L^2}{16 \sqrt{x y (L-x) (L-y)}}\int du\,dv\, u\, \text{csch}(\pi  u) \text{csch}(\pi  v) \left(\left(H_{-\frac{1}{2} i (u+v-i)}+H_{\frac{1}{2} i (u+v+i)}\right) (\sinh (\pi  u)-\sinh (\pi
   v))\right.\nn\\&&\left.-\left(H_{-\frac{1}{2} i (u-v-i)}+H_{\frac{1}{2} i (u-v+i)}\right) (\sinh (\pi  u)+\sinh (\pi  v))\right) e^{-\frac{1}{2} i (u
   (z(x)-z(y))+v (z(x)+z(y)))}\label{integrand}.
\eea
The singular terms come from the constant values of the factor of the exponential in the integrand in the limit $u,v\rightarrow \pm \infty$.  Subtracting these constant limits of \eqref{integrand} we obtain an integral in $u, v$ that is finite, but difficult to compute. This will lead to a completely non local contribution. However, our interest is to compute the contribution of singular terms to the modular Hamiltonian. In order to extract the term proportional to $\delta(x-y)$ we use that the expansion of the integrand when $u\rightarrow\infty$ goes as $4\,v\cosh(v)$. Performing the integrals in $u$ and $v$ we obtain the  result from \eqref{integrand}
\be
\frac{2 \pi  m\, y (L-y)}{L}\delta(x-y).
\ee

Now, we can analyse the $v\rightarrow\infty$ limit of the integrand in \eqref{integrand} in order to extract another singular term for the modular Hamiltonian coming from this limit. The resulting expression is
\be
\frac{m L^2}{4 \sqrt{x y (L-x) (L-y)}}\int du\,dv\, u\, \text{csch}(\pi  u) \left(\log \left(\frac{1}{|v|}\right)-\gamma_E +\log (2)\right)e^{-\frac{1}{2} i (u
   (z(x)-z(y))+v (z(x)+z(y)))}\,.
\ee
 The term proportional to $(-\gamma_E+\log(2))$ can be integrated and gives
\be
2\pi m (-\gamma_E+\log(2))\frac{x y}{L}\delta(x+y-L).
\ee
The integral of the term involving the $\log\left(\frac{1}{|v|}\right)$ leads to
\be
-\frac{\pi\,m \,L^2 }{\left(\sqrt{x} \sqrt{L-y}+\sqrt{y} \sqrt{L-x}\right)^2 |\log \left(\frac{x y}{(L-x)(L-y)}\right)|}-2\pi\,m\,\gamma_E\frac{x\,y}{L}\delta(x+y-L).
\ee

Then, the final result for the singular terms of the first order correction to the Modular Hamiltonian can be written
\bea
H_{1,0}+H_{1,1}&=&2 \pi m \frac{  \, x (L-x)}{L} \delta(x-y)\gamma^0\nn\\&& + 2\, \pi m \frac{  \, x (L-x)}{L} \left(2\log2-3\gamma_E-\log(m L)\right) \delta(x+y-L)\gamma^0  +\nn\\&& +\,\frac{m\pi\,x(L-x)}{L|y-L+x|}\gamma^0+ \textrm{non singular.} \label{result}
\eea
We recognize the local term is equivalent to the term that would come from (\ref{espre}) if we use the massive expression for $T_{00}$. We see a term like the last one in (\ref{result}) is seen numerically in figure \ref{h01:c}.

\section{Propagation and subleading local terms for a Dirac field in $d=2$}
\label{ap2}

 The modular Hamiltonian written in a spatial surface $\Sigma$ is
\be
K=\int_{\Sigma} ds_1\,ds_2\, \psi^\dagger(s_1) H(s_1,s_2) \psi(s_2)
\ee
where $s_1,s_2$ are distance parameters.

The modular Hamiltonian can be written in any other surface for the same causal region $V$ using the propagation equation for the field
\bea
\psi(x)=\int_{\Sigma^\prime} ds^\prime\, S(x-x^\prime)\gamma^\mu\eta^\prime_\mu(s^\prime) \psi(s^\prime)\,,\label{54}\\
\bar{\psi}(x)=\int_{\Sigma^\prime} ds^\prime\, \bar{\psi}(s^\prime) \gamma^\mu\eta^\prime_\mu(s^\prime) S(x^\prime-x)\,,
\eea
where
\be
S(x-y)=\{\psi(x),\bar{\psi}(y)\}=  (i\gamma^\mu \partial_\mu+m) i\Delta(x-y)  \,,
\ee
and $\eta^\mu$ is the normal to the surface.
We are using signature $(+,-,-,\hdots)$.

Then let us look at how the local terms propagate. A local term in $\Sigma^\prime$ will arise from a local term in $\Sigma$ convoluted with the singular terms in the Green function $S(x-x^\prime)$. Because the integrals that give place to a local term in $\Sigma^\prime$ are for a compact region of the coordinate $s$, no other singular terms can arise from the finite terms. Let us then look at the singularity structure of the Green function.
We have
\be
i\Delta(x)=[\phi(x),\phi(0)]=\int \frac{d^2p}{2\pi}\,\epsilon(p^0) \delta(p^2-m^2) e^{-i p x} =\epsilon(x^0) \theta(x^2) i \frac{\textrm{Im}(K_0(i m \sqrt{x^2}))}{\pi}\,.
\ee
Since
\be
i\frac{\textrm{Im}(K_0(i y))}{\pi}\sim -\frac{i}{2} + {\cal O}(y^2)
\ee
for small positive argument $y$, we have that $i\Delta(x)$ is smooth everywhere except at the null cone where it has a jump which is the same as the one of the function
\be
-\frac{i}{4} (\epsilon(x^+)+\epsilon(x^-))\,.
\ee
This is constant inside each of the two light cones.

We will be using null coordinates
\bea
x^+&=&x^0+x^1\,,\hspace{2.5cm}
x^-=x^0-x^1\,,\\
x^0&=&(x^++x^-)/2\,,\hspace{1.7cm}
x^1=(x^+-x^-)/2\,,\\
\partial_+&=&\frac{1}{2}(\partial_0+\partial_1)\,,\hspace{2cm}\partial_-=\frac{1}{2}(\partial_0-\partial_1)\,, \\
g^{\mu\nu}&=&\left(\begin{array}{cc} 0 & 2 \\ 2 & 0  \end{array}\right)\,,\hspace{2cm}  g_{\mu\nu}=\left(\begin{array}{cc} 0 & 1/2 \\ 1/2 & 0  \end{array}\right)\,.
\eea
We will also be using the chiral representation for Dirac matrices where $\gamma^3=\gamma^0\gamma^1$ is diagonal
\be
\gamma^3=\left(\begin{array}{cc} 1 & 0 \\ 0 & -1  \end{array}\right)\,,\hspace{2cm}
\gamma^0=\left(\begin{array}{cc} 0 & 1\\ 1 & 0  \end{array}\right)\,.
\ee
The projection over chiralities are
\be
Q^+=\frac{1+\gamma^3}{2}\,,\hspace{2cm}Q^-=\frac{1-\gamma^3}{2}\,,
\ee
and
\be
\gamma^\mu \partial_\mu=2\gamma^0(Q^+\partial_++ Q^-\partial_-)\,.
\ee

The singularity structure of the anticommutator in chiral representation and null coordinates is then given by
\bea
S(x)\simeq\frac{1}{4}(\gamma^\mu \partial_\mu-i m)(\epsilon(x^+)+\epsilon(x^-))=\frac{1}{4}(2\gamma^0(Q^+\partial_++ Q^-\partial_-)-i m)(\epsilon(x^+)+\epsilon(x^-))\nonumber\\
=\gamma^0  \left(\begin{array}{cc}
\delta(x^+) & -i \frac{m}{4} (\epsilon(x^+)+\epsilon(x^-))\\-i \frac{m}{4} (\epsilon(x^+)+\epsilon(x^-)) &  \delta(x^-)
\end{array}\right)\,,\label{gf}
\eea
plus less singular terms.

Let us think in the propagation of a local term proportional to the stress tensor
\be
\int_{\Sigma} ds\, \eta_\mu T^{\mu\nu} a_\nu \label{intre}\,.
\ee
The stress tensor writes (both for massive and massless fields)
\be
T^{\mu\nu}=\frac{i}{4}\bar{\psi}(\gamma^\mu \stackrel{\leftrightarrow}{\partial}^\nu+\gamma^\nu \stackrel{\leftrightarrow}{\partial}^\mu)\psi\label{ten}\,.
\ee
We are interested in the propagation from the operator in a given point $x$ on the surface $\Sigma$ to another surface $\Sigma^\prime$. Without loss of generality we can take the coordinate system such that the normal vector $\eta_\mu(x)=(1,0,\hdots)$.    The operator at $x$ in the integral (\ref{intre}) then writes
\bea
a^0 T_{00}+a^1 T_{01}=\bar{\psi}(x) (a^0 (-\frac{i}{2}) \gamma^1 \stackrel{\leftrightarrow}{\partial}_x + \frac{i}{2} a^1 \gamma^0 \stackrel{\leftrightarrow}{\partial}_x) ) \psi(x)\nonumber\\
=\frac{i}{2}\bar{\psi}(x)\gamma^0 (a^+ Q_- -a^- Q_+ ) \stackrel{\leftrightarrow}{\partial}_x \psi(x)\label{esa}
\,.\eea
Here the symmetrized derivatives act only on the fermion fields and not on the components of $a^\mu$. The propagation is given by
\be
\int_{\Sigma^\prime} ds_1^\prime \, ds_2^\prime\,\int_\Sigma dx\, \bar{\psi}(x_1^\prime)\eta_\beta(s_1^\prime)\gamma^\beta S(x_1^\prime-x) \frac{i}{2}\gamma^0 (a^+ Q_- -a^- Q_+ ) \stackrel{\leftrightarrow}{\partial}_x  S(x-x_2^\prime) \eta_\alpha(s_2^\prime)\gamma^\alpha\psi(x_2^\prime)\label{hjhk}\,.
\ee
Using that for the integrals on the length parameter along a surface
\be
\gamma^0 \eta_\alpha(s)\gamma^\alpha ds=\left(\begin{array}{cc} - dx^- & 0 \\ 0 & dx^+ \end{array}\right)\,,
\ee
we can rewrite (\ref{hjhk}) as
\bea
\int_{\Sigma^\prime} \int_\Sigma dx\, \bar{\psi}(x_1^\prime)\gamma^0 \left(\begin{array}{cc} - dx_1^{\prime -} & 0 \\ 0 & dx_1^{\prime +} \end{array}\right) \gamma^0  \left(\begin{array}{cc}
\delta(x^{\prime+}_1-x^+) & 0\\0 &  \delta(x_1^{\prime -}-x^-)
\end{array}\right) \frac{i}{2}\gamma^0 (a^+ Q_- -a^- Q_+ ) \stackrel{\leftrightarrow}{\partial}_x  \nonumber\\
\times\,\,\,\,\,  \gamma^0  \left(\begin{array}{cc}
\delta(x^+-x^{\prime+}_2) & 0\\0 &  \delta(x^--x_2^{\prime -})
\end{array}\right) \gamma^0 \left(\begin{array}{cc} - dx_2^{\prime -} & 0 \\ 0 & dx_2^{\prime +} \end{array}\right)\psi(x_2^\prime)\nonumber \\
=-\int_{\Sigma^\prime}dx^{-} a^- (x^-)\bar{\psi}(x^-)\gamma^0Q^+ \frac{i}{2}\stackrel{\leftrightarrow}{\partial}_{x^-}\psi(x^-)+\int_{\Sigma^\prime}dx^{+} a^+ (x^+)\bar{\psi}(x^+)\gamma^0Q^- \frac{i}{2}\stackrel{\leftrightarrow}{\partial}_{x^+}\psi(x^+)\,,\label{ope}
\eea
where $a^{\pm}$ is computed on the surface $\Sigma$. Then if $a^\mu$ is defined everywhere by the components $a^-(x^-)$, independently of $x^+$, and $a^+(x^+)$, independent of $x^-$, as given by the values on $\Sigma$, we have using (\ref{ten}) on $\Sigma^\prime$ that the operator (\ref{ope}) reads
\be
\int_{\Sigma^\prime} ds \,\eta^\mu a^\nu T_{\mu\nu}=\int_{\Sigma^\prime} ds \,(\eta^+ a^+ T_{++}+\eta^- a^- T_{--})=\int_{\Sigma^\prime} dx^+ \, a^+ T_{++}-\int_{\Sigma^\prime} dx^- \, a^- T_{--} \,,
\ee
where we have used
\be
ds (\eta^+,\eta^-)=(dx^+,-dx^-)\,.
\ee
We have changed the signs due to the differentials $x^-$ with respect to $x$ but we have not changed the integration limits.\footnote{Note that this practice is convenient because we only have to keep track of the change on the differentials but it involves a subtle point: it requires we introduce a sign when a delta function on the $x^-$ variable has been eliminated by integration.}
This keeps the same form as in the original surface. In this case no non local terms are generated.

Of course, for the massless field this is just a consequence of the fact that $j^\mu=a_\nu T^{\mu\nu}$ is a conserved current and its flux is independent of the surface. In fact when $\partial_+ a^-=\partial_- a^+=0$, using that in the massless case the trace is zero $T^{+-}=0$, and $\partial_+ T^{++}=0$, $\partial_- T^{--}=0$, by conservation,
\be
\partial_\mu T^{\mu\nu}a_\nu=T^{++}\partial_+ a_++T^{--}\partial_- a_-=0\,.
\ee

In the massive case we have two changes. The first one is the change in the equation (\ref{esa}) to incorporate the mass term in the stress tensor
\be
a^0 T_{00}+a^1 T_{01}=
=\frac{i}{2}\bar{\psi}(x)\gamma^0 (a^+ Q_- -a^- Q_+ ) \stackrel{\leftrightarrow}{\partial}_x \psi(x)+ a^0 \bar{\psi}(x) m \psi(x)\,.
\ee
The second change is the additional terms in the propagator. These are non singular, but contracted with the derivative term of the stress tensor kernel will produce delta functions. Only first order terms in the mass will produce delta functions, all combinations with more powers give non localized terms. The term produced by the addition to the stress tensor gives
\bea
\int_{\Sigma^\prime} \int_\Sigma dx\, \bar{\psi}(x_1^\prime)\gamma^0 \left(\begin{array}{cc} - dx_1^{\prime -} & 0 \\ 0 & dx_1^{\prime +} \end{array}\right) \gamma^0  \left(\begin{array}{cc}
\delta(x^{\prime+}_1-x^+) & 0\\0 &  \delta(x_1^{\prime -}-x^-)
\end{array}\right)  a^0 m \nonumber  \\
\times\,\,\,\,\,  \gamma^0  \left(\begin{array}{cc}
\delta(x^+-x^{\prime+}_2) & 0\\0 &  \delta(x^--x_2^{\prime -})
\end{array}\right) \gamma^0 \left(\begin{array}{cc} - dx_2^{\prime -} & 0 \\ 0 & dx_2^{\prime +} \end{array}\right)\psi(x_2^\prime)\,.
\label{ope1}
\eea
Note this term does not produce localized terms in the surface $\Sigma^\prime$ since it involves a delta function between two points in $\Sigma^\prime$ whose coordinates $x^+$ and $x^-$ coincide with the corresponding coordinate of the point $x$ in $\Sigma$. That is, the singular perturbations propagate at light velocity in different directions generating a possible quasilocal term.

Then it rests the perturbations of the propagator computed with the massless kernel for the stress tensor. We have
\bea
\int_{\Sigma^\prime} \int_\Sigma dx\, \bar{\psi}(x_1^\prime)\gamma^0 \left(\begin{array}{cc} - dx_1^{\prime -} & 0 \\ 0 & dx_1^{\prime +} \end{array}\right) \gamma^0  \left(\begin{array}{cc}
\delta(x^{\prime+}_1-x^+) & 0\\0 &  \delta(x_1^{\prime -}-x^-)
\end{array}\right) \frac{i}{2}\gamma^0 (a^+ Q_- -a^- Q_+ ) \stackrel{\leftrightarrow}{\partial}_x \nonumber\\
\times\,\,\,\,\,  -i \frac{m}{4} (\epsilon(x^+-x^{\prime+}_2)+\epsilon(x^--x^{\prime-}_2)) \gamma^0 \left(\begin{array}{cc} - dx_2^{\prime -} & 0 \\ 0 & dx_2^{\prime +} \end{array}\right)\psi(x_2^\prime)+h.c.\label{ope2}
\eea
The two derivatives with different directions can be made to act to the same side since derivating the components of $a^\mu$ will produce non singular terms. We get
\bea
\int_{\Sigma^\prime} \int_\Sigma dx\, \bar{\psi}(x_1^\prime)\gamma^0 \left(\begin{array}{cc} - dx_1^{\prime -} & 0 \\ 0 & dx_1^{\prime +} \end{array}\right) \gamma^0  \left(\begin{array}{cc}
\delta(x^{\prime+}_1-x^+) & 0\\0 &  \delta(x_1^{\prime -}-x^-)
\end{array}\right) \gamma^0 \frac{1}{2}(a^+ Q_- -a^- Q_+ )  \nonumber \\
\times\,\,\,\,\,   m (\delta(x^+-x^{\prime+}_2)-\delta(x^--x^{\prime-}_2)) \gamma^0 \left(\begin{array}{cc} - dx_2^{\prime -} & 0 \\ 0 & dx_2^{\prime +} \end{array}\right)\psi(x_2^\prime)+h.c.\label{ope3}
\eea
This has local and quasilocal terms. The quasilocal terms  cancel the ones produced by (\ref{ope1}).

The local terms are
\bea
-\frac{m}{2}\int_{\Sigma^\prime} \psi^\dagger(x)  \gamma^0    \psi(x) (-a^-(x^-) dx^{+} +a^+(x^+) dx^-)\,.
\eea
Using
\be
-a^- dx^++a^+ dx^-=-2 ds a_\mu \eta^\mu\,,
\ee
we get
\be
m\int_{\Sigma^\prime} ds\, \psi^\dagger(x)  \gamma^0    \psi(x) \, a_\mu \eta^\mu\,.
\ee
That is, the local term generated on $\Sigma^\prime$ by the original term in $\Sigma$ has again the form of the flux of the stress tensor where the vector $a^\mu$ has propagated in the same way than in the massless case. The mass appearing in the new surface is the field mass coming from the propagator, disregarding the possible mass term we could have written in the original surface. The term in derivatives generates the mass term by itself, with the propagating mass.

\end{document}